\documentclass[12pt]{article}
\usepackage{amssymb}
\usepackage{comment}
\usepackage{amsmath}
\usepackage{amsthm}
\usepackage{mathrsfs}
\usepackage{adjustbox}
\usepackage{graphicx}
\usepackage{listings}
\usepackage{bm}
\usepackage{subfig}
\usepackage{booktabs}
\usepackage{dsfont}
\usepackage{tikz}
\usepackage{stmaryrd}
\usepackage{mathtools}
\usepackage[hidelinks]{hyperref}
\usepackage[round]{natbib}
\usepackage{xcolor}
\usepackage[T1]{fontenc}
\usepackage{stackengine}
\usepackage{algorithm}
\SetSymbolFont{stmry}{bold}{U}{stmry}{m}{n}
\usepackage{algpseudocode, algorithm, algorithmicx}
\usepackage{graphicx}
\usepackage{psfrag,epsf}
\usepackage{enumerate}
\usepackage{natbib}
\usepackage{url} 

\newcommand{\blind}{0}

\newcommand{\esp}[1]{\mathbb{E}\left[#1\right]}
\newcommand{\var}[1]{\mathrm{Var} \left[#1\right]}

\newcommand{\one}[1]{\mathds{1}_{#1}}
\newcommand{\figheight}{100}
\newcommand{\figwidth}{100}

\newcommand*\Let[2]{\State #1 $\gets$ #2}
\newcommand*\Compute[2]{\State #1 for #2}


\theoremstyle{plain}

\begin{document}

\def\spacingset#1{\renewcommand{\baselinestretch}%
{#1}\small\normalsize} \spacingset{1}


\if0\blind
{
  \title{\bf Spline Regression with Automatic Knot Selection}
  \author{Vivien Goepp
		  \thanks{
	\texttt{vivien.goepp@parisdescartes.fr}}\hspace{.2cm}\\
    MAP5 (CNRS UMR 8145), Universit\'e Paris Descartes\\
	Olivier Bouaziz\\
    MAP5 (CNRS UMR 8145), Universit\'e Paris Descartes,\\
    and \\
    Gr\'egory Nuel \\
    LPSM (CNRS UMR 8001), Sorbonne Universit\'e}
  \maketitle
} \fi

\if1\blind
{
  \bigskip
  \bigskip
  \bigskip
  \begin{center}
    {\LARGE\bf Title}
\end{center}
  \medskip
} \fi

\bigskip
\begin{abstract}
In this paper we introduce a new method for automatically selecting knots in spline regression.
The approach consists in setting a large number of initial knots and fitting the spline regression through a penalized likelihood procedure called adaptive ridge.
The proposed method is similar to penalized spline regression methods (e.g. P-splines), with the noticeable difference that the output is a sparse spline regression with a small number of knots.
We show that our method -- called A-spline, for \emph{adaptive splines} -- yields sparse regression models with high interpretability, while having similar predictive performance similar to penalized spline regression methods.
A-spline is applied both to simulated and real dataset.
A fast and publicly available implementation in \texttt{R} is provided along with this paper.
%
\end{abstract}
\noindent%
\vfill

\newpage
\spacingset{1.45} 
\section{Introduction}
\label{section:introduction}

Spline regression has known a great development in the past decades \citep[see][]{Wahba1990SplineModelsObservational, Hastie2001ElementsStatisticalLearning, Ruppert2009Semiparametricregression2003, Wood2017GeneralizedAdditiveModelsa} and has become a tool of choice for semiparametric regression.
This success can be explained by the fact that splines are restrictive enough to benefit from the simplicity of parametric estimation, and yet are general enough to accurately approximate a large variety of smooth function.
Spline regression is performed by choosing a set of knots and by finding the spline defined over these knots that minimizes the residual sum of squares.
The number of knots has an important influence in the resulting fit: with not enough knots the regression is underfitted and with too many knots it is overfitted.
Choosing the position of knots is also an issue since uniformly distributed knots can lead to overfitting in an area where there are few points and underfitting in an area where there are many points.

The most widely used spline regression methods overcome this difficulties by using a penalization approach.
In smoothing splines, knots are set at each data point and the wiggliness of the spline is controlled by penalizing over its integrated squared second order derivative $\int_{}^{}\left\{ f''\left( t \right) \right\}^{2}dt$.
The smoothing spline estimate has a closed-form expression and computationally efficient techniques have been developed.
We refer to \citep[][Section~5]{Hastie2001ElementsStatisticalLearning} for a detailed explanation on smoothing splines.
\cite{OSullivan1986StatisticalPerspectiveIllPosed} generalized smoothing splines to an arbitrary choice of knots. This allows to set fewer knots than the sample size.
Two \texttt{R} implementations are available in the package \texttt{gam} \citep{Hastie2018gamGeneralizedAdditive, Hastie2001ElementsStatisticalLearning} and the package \texttt{mgcv} \citep{Wood2017GeneralizedAdditiveModelsa}.
Later, \cite{Eilers1996FlexibleSmoothingBsplines, Marx1998Directgeneralizedadditive} introduced a penalty based on the finite order differences of the parameters.
The corresponding splines are called P-splines.
This penalization is closely related to that of O'Sullivan \citep[see][Section 3]{Eilers1996FlexibleSmoothingBsplines}: it is simpler since no integration is involved, and it allows for generalizations to derivatives of higher order.
However, O-Sullivan's penalty is more general in that the knots do not have to be equally spaced.
See \citet{Wand2010SemiparametricRegressionSullivan} and \citet[][Appendix A]{Eilers2015TwentyyearsPsplines} for comparisons of the two methods.
A detailed review of P-splines is given in \citet{Eilers2015TwentyyearsPsplines} and citations therein.
We note that P-splines are also closely related to \cite{Whittaker1922NewMethodGraduation}'s graduation method, which can be seen as a P-spline of order $0$ with knots placed at data points.

These regularized approaches in spline regression are simple and computationally fast.
However, a spline regression with fewer knots is easier to interpret, which in many cases is a desired goal.
Thus, some attempts have been made to find a non-penalized regression procedure with an automatic selection of knots.
The idea is to choose more knots -- and so basis splines -- in data-dense regions where the underlying function has more variability.
One could try to find the best knots by setting a very large number of knots and exploring the set of splines defined on any subset of the knots.
But as pointed out by \cite{Wand2000ComparisonRegressionSpline}, this method is not tractable in practice.
Previous attempts to find the best number and location of knots can be found in the literature; we refer to \citet{Wand2000ComparisonRegressionSpline} for a review.
\cite{Friedman1991MultivariateAdaptiveRegression} has developed a multivariate variable selection technique called MARS (Multivariate Adaptive Regression Splines).
It uses a recursive partitioning of the domain and sequentially selects the most relevant knots with a forward step size procedure followed by backward step size procedure.
See also \citep{Friedman1989FlexibleParsimoniousSmoothing} and \citep[][Section 9.4]{Hastie2001ElementsStatisticalLearning} for details.
\cite{Luo1997HybridAdaptiveSplines} have later developed a closely related approach for automatic selection of knots called Hybrid Adaptive Splines.
Like MARS, it uses a forward stepwise regression procedure and instead of using a backward procedure to remove unnecessary knots, it fits penalized splines.
Other paths have been taken to solve this computationally intensive problem.
Namely, \cite{Jamrozik2010SelectionLocationsKnots} have offered to estimate the best location of knots using a differential evolution algorithm.
However, their approach was limited to a number of knots varying between $4$ and $7$ and to splines of order $1$.

In this article, we introduce a new computationally efficient method to automatically select the number and position of the knots from the data.
It is called A-splines, for \emph{adaptive splines}.
It is based on a regularization method with an approximate L$_{0}$ norm penalty.
Although our approach is different from P-splines, A-spline regression uses an objective function closely related to that of P-spline.
Our method is defined for splines of any order $q \geq 0$.
In particular, using splines of order $0$ -- i.e piecewise constant functions -- allows to perform automatic detection of breakpoints.
Splines of order $1$, i.e. continuous broken lines, can be used as a generalization of the linear model which allows for shifts in the slope.
In most cases when the true function $f$ is assumed to be ``smooth'', splines of order $3$ are used, which yield a sparser model than the state-of-the-art spline regression methods.
Therefore, our method is to be preferred when the simplicity of the model is a desired feature.

This paper is constructed as follows. Section~\ref{section:spline} gives a short summary of B-splines and B-spline regression.
Section~\ref{section:a_spline} introduces our spline regression method.
In Section~\ref{section:glm}, our method is extended to the generalized linear model framework.
Section~\ref{section:penalty} deals with the choice of the bias-variance tradeoff parameter.
Section~\ref{section:simu} compares the prediction performance of our model to P-splines through a simulation study.
Section~\ref{section:implementation} gives some details about the fast implementation of the fitting algorithm.
Finally, A-spline is illustrated on several real datasets in Section \ref{section:real_data}.

\section{B-spline Regression}
\label{section:spline}
\subsection{B-spline Basis}
\label{section:b_spline}

\begin{figure}
	\centering
    \subfloat[Order $0$ B-splines]{
	  \includegraphics[width = 0.45\textwidth]{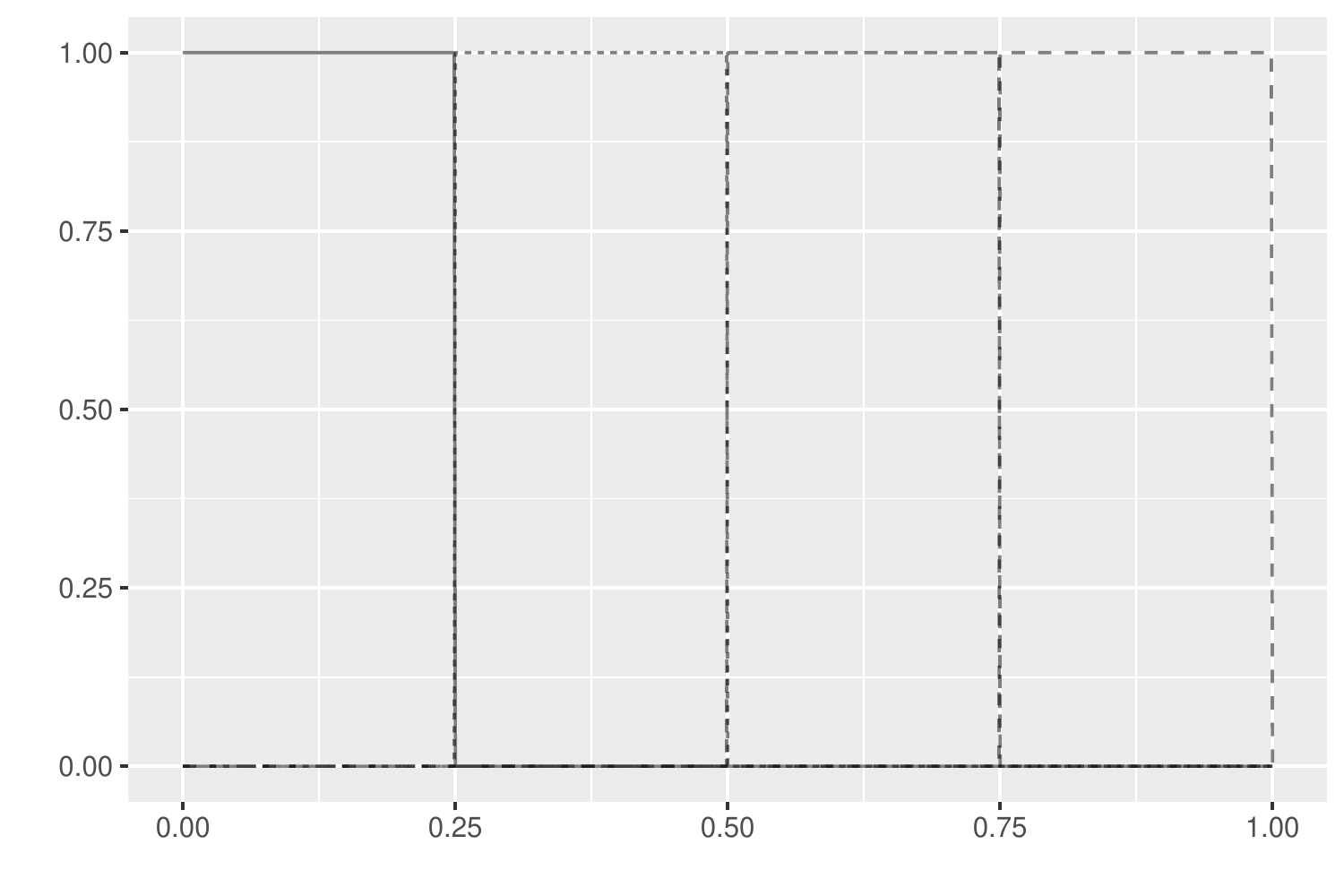}
			\label{fig:b_spline_0}
	}
	\hfil
    \subfloat[Order $1$ B-splines]{
	  \includegraphics[width = 0.45\textwidth]{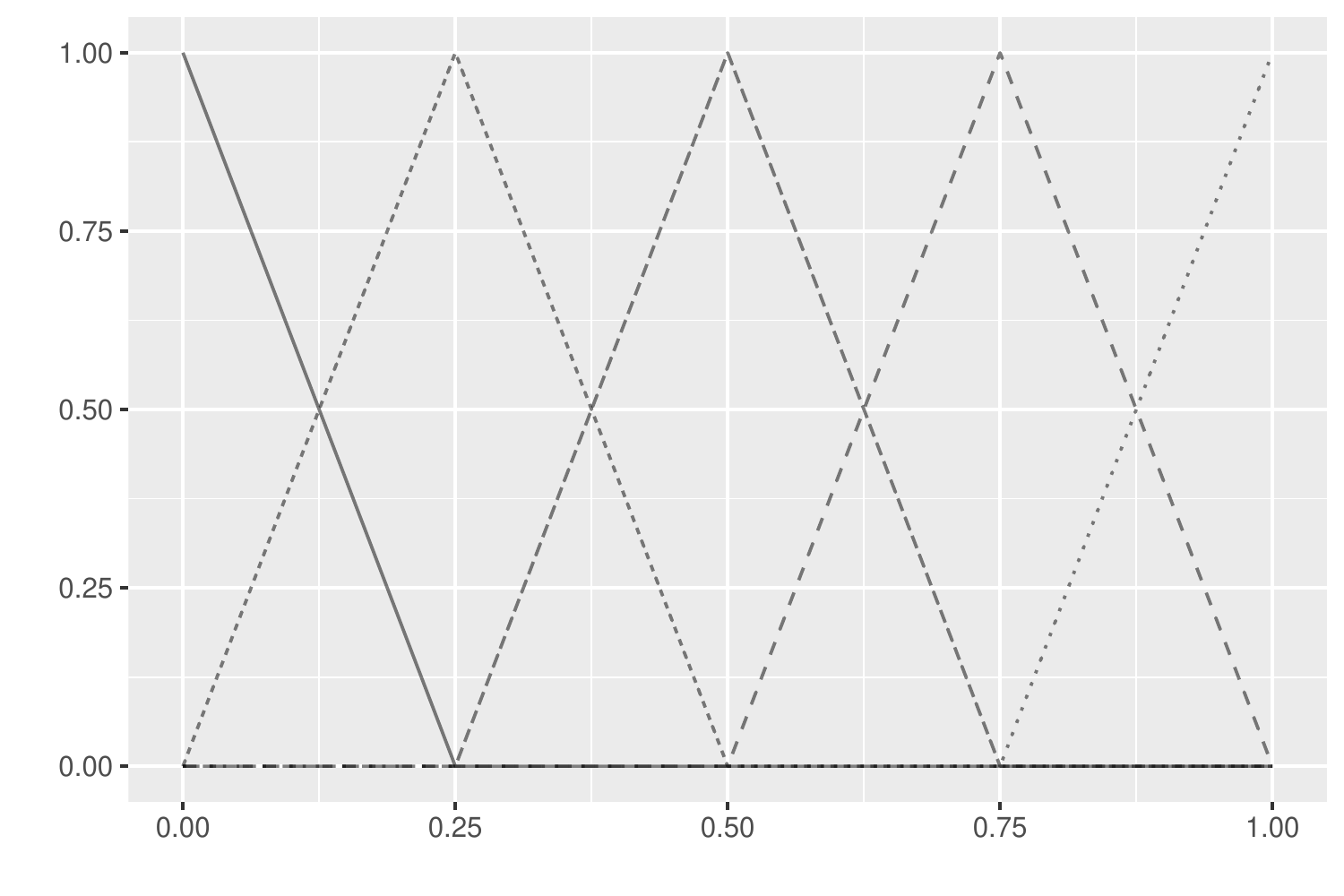}
			\label{fig:b_spline_1}
	}
	\hfil
    \subfloat[Order $2$ B-splines]{
	  \includegraphics[width = 0.45\textwidth]{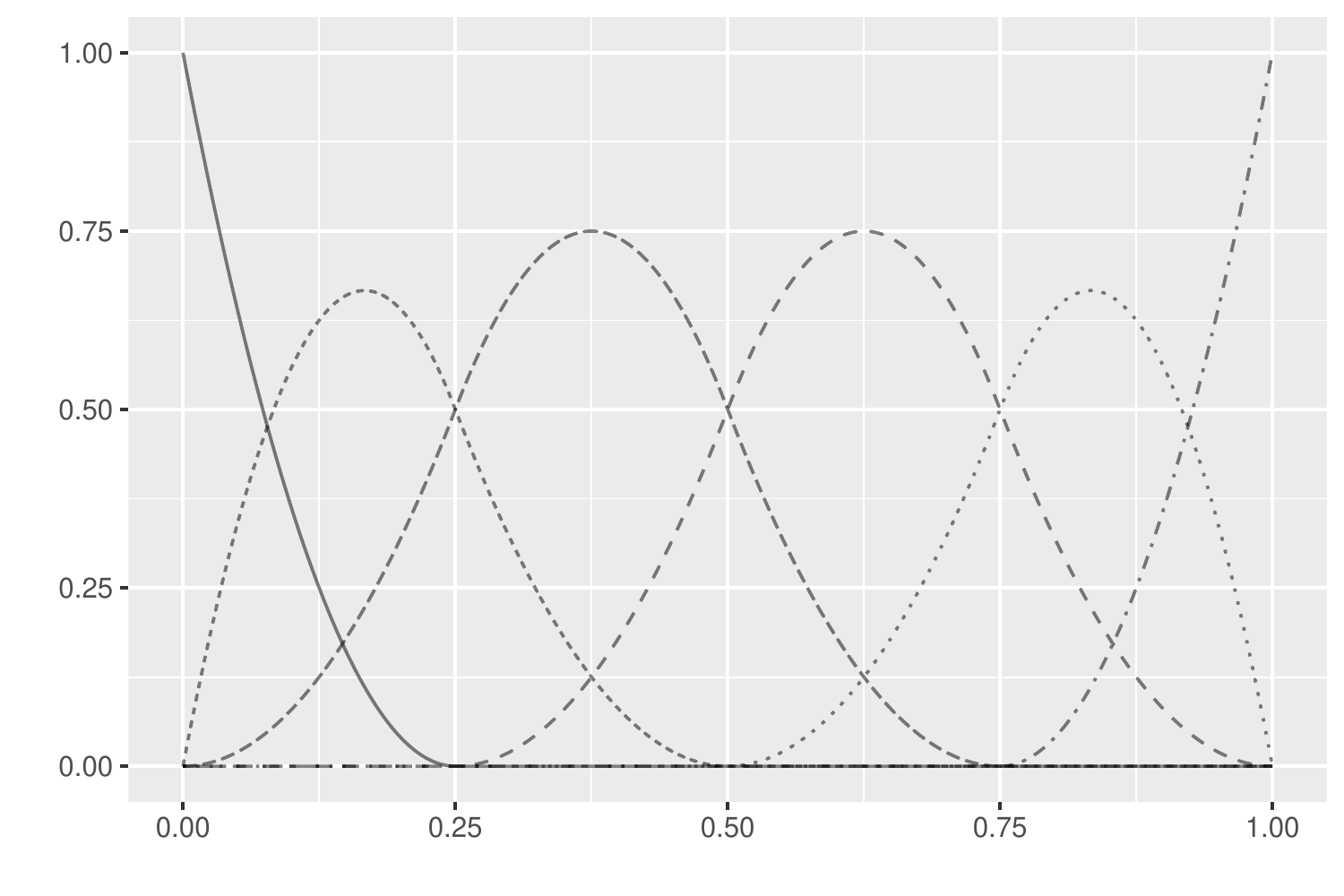}
			\label{fig:b_spline_2}
	}
	\hfil
	\subfloat[Order $3$ B-splines]{
	  \includegraphics[width = 0.45\textwidth]{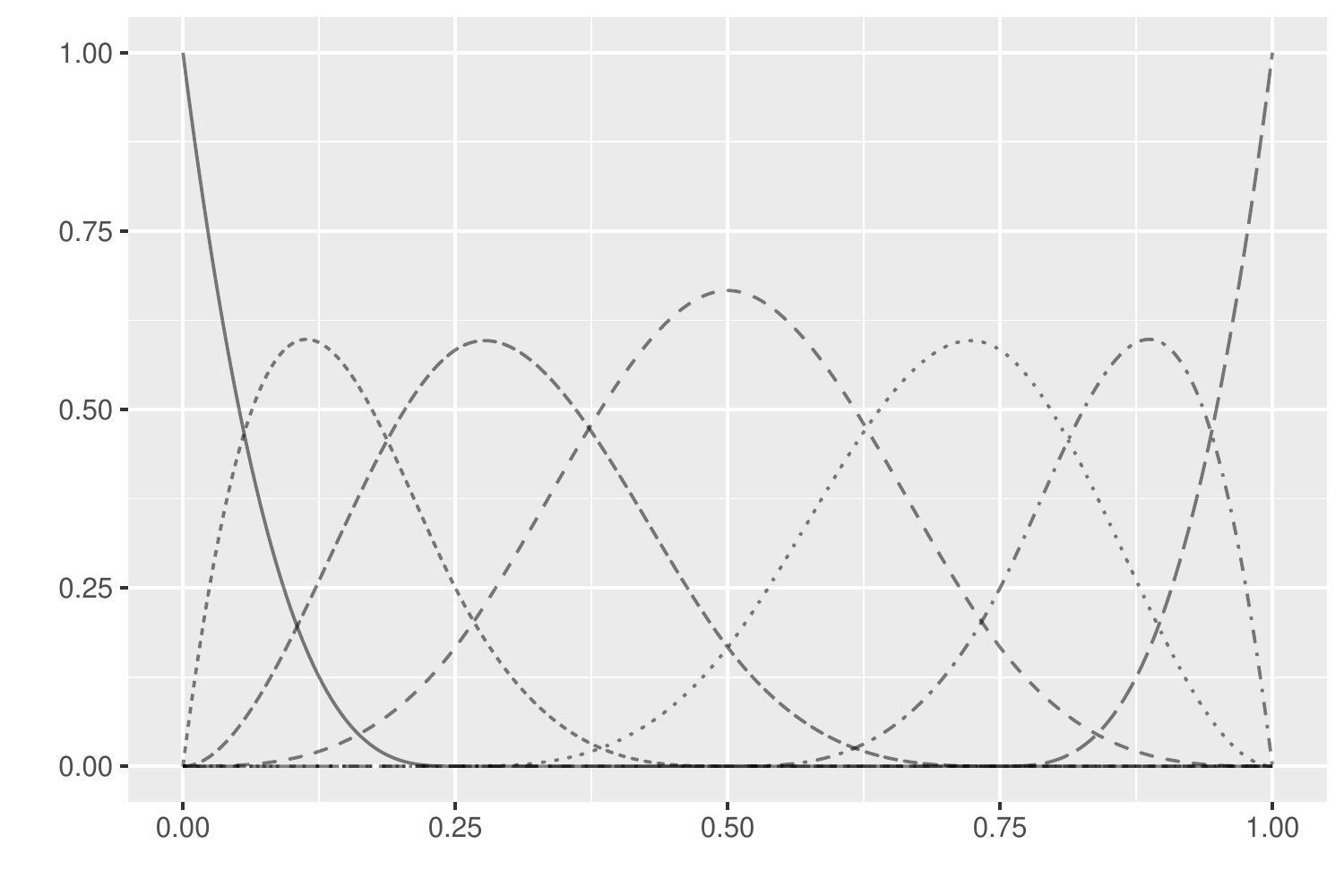}
			\label{fig:b_spline_3}
	}
	\hfil
	\caption{Bases of B-spline of order $0$ to $3$ (Panels a to d) with $3$ knots: $\left( 0.25, 0.5, 0.75 \right)$. Note that with $3$ knots, there are $4$ splines in the basis of order $0$ and $7$ splines in the basis of order $3$.}
	\label{fig:b_spline}
\end{figure}
	  
In this section we recall the definition and some basic properties of splines and B-splines.
Throughout this work, let $t_{1}, \dots, t_{k}$ be the ordered knots included in a real interval $\left[ a, b \right]$.
A spline of order $q \geq 0$ is a piecewise polynomial function of order $q$ such that its derivatives up to order $q - 1$ are continuous at every knot $t_{1}, \dots, t_{k}$.
The set of splines of order $q$ over the knots $\bm{t} = \left( t_{1}, \dots, t_{k} \right)$ is a vector space of dimension $q + k + 1$.

A possible choice of spline basis is the truncated power basis: $\{x^{0}$, $\dots, x^{q}, (x - t_{1})_{+}^{q}, \dots$, $\left( x - t_{k} \right)_{+} ^{q}\}$, where $\left( u \right)_{+} = \max \left( u, 0 \right)$.
The first $q + 1$ functions of the basis are polynomials and the other $k$ functions are truncated polynomials of degree $q$.
Decomposing a spline into the truncated power basis brings out powers of large numbers, which lead to rounding errors and numerical inaccuracies \citep[][p. 85]{DeBoor1978PracticalGuideSplines}.

In order to solve this problem, \cite{DeBoor1978PracticalGuideSplines} introduced a spline basis -- called B-splines -- more adapted to computational implementation of spline regression.
A B-spline is a spline which is non-zero over $[x_{k}, x_{k + q + 1}]$ for some $k$.
For $i = 1, \dots, q + k + 1$, the $i$-th B-spline of order $q$ is noted $B_{i, q}\left( x \right)$ and is defined by
\begin{equation*}
		B_{i, q}\left( x \right) = \frac{x - t_{i}}{t_{i + q} - t_{i}} B_{i, q - 1}\left( x \right) + \frac{t_{i + q + 1} - x}{t_{i + q + 1} - t_{i + 1}} B_{i + 1, q + 1}\left( x \right) \quad \text{if} \quad q > 0
\end{equation*}
and $B_{i, 0}\left( x \right) = \one{t_{i} \leq x < t_{i + 1}}$.
Important properties of a B-spline are: (i) the B-spline is non-zero over an interval spanning $q + 2$ knots; (ii) at a point, only $q + 1$ B-splines are non-zero; (iii) $B_{i, q}\left( x \right) \in [0, 1]$.
An illustration of B-spline bases of order $0$ to $3$ is given in Figure \ref{fig:b_spline}.
In practice, B-splines can be computed using the function \texttt{bSpline} from the \texttt{R} package \texttt{splines2} \citep{Wang2017splines2RegressionSpline}.

\subsection{B-Spline Regression}
\label{section:b_spline_regression}
Let $\left( x_{i}, y_{i} \right) \in \mathbb{R} \times \mathbb{R}$ be the univariate data and consider the non-parametric regression  setting
\begin{equation}
		y_{i} = f\left( x_{i} \right) + \varepsilon_{i}, \quad 1 \leq i \leq n,
		\label{eq:regression}
\end{equation}
with i.i.d. Gaussian errors $\varepsilon_{i}$ and where $f$ is a ``smooth'' function.
The function $f$ is estimated by a spline over an interval $[a, b]$ containing all $x_{i}$s.
Fitting the data consists in minimizing the sum of squares
\begin{equation}
		\mathrm{SS}\left( \bm{a} , \bm{t}\right) = \sum_{i = 1}^{n} \left\{ y_{i} - \sum_{j = 1}^{q + k + 1} a_{j} B_{j, q}\left( x_{i} \right) \right\}^{2},
		\label{eq:sum_square}
\end{equation}
where $\bm{a} = \left( a_{1}, \dots, a_{q + k + 1} \right)$ is the B-spline coefficients. 
The knots $\bm{t}$ are present as parameter of $\mathrm{SS}$ to highlight that the whole fitting procedure depends on the choice of the knots.
This is the framework of ordinary least squares regression with design matrix $\bm{B} = \left[ B_{j, q}\left( x_{i} \right) \right]_{i, j}$ and parameter $\bm{a}$:
\begin{equation}
		\mathrm{SS}\left( \bm{a}, \bm{t} \right) = \|\bm{y} - \bm{B}\bm{a}\|_{2}^{2}.
		\label{eq:sum_square_2}
\end{equation}

\section{Automatic Selection of Knots}
\label{section:a_spline}

When there are many knots, spline regression is prone to overfitting.
In the extreme case, when there as as many parameters as data points, the fitted spline interpolates the data.
In this paper, we propose to estimate the spline which makes the best tradeoff between model dimension (i.e. number of knots) and goodness of fit.
To this effect, we choose a high number of equally spaced initial knots and penalize over the number of knots.
When a B-spline is defined over the knots $t_{1}, \dots, t_{k}$ and is such that $\Delta^{q + 1}a_{j^{*}} = 0$ for some $j^{*}$, it can be reparametrized as a B-spline over the knots $t_{1}, \dots, t_{j^{*} - 1}, t_{j^{*} + 1}, \dots t_{k}$.
Consequently, one would like to penalize over the number of non-zero $q + 1$-order differences:
\begin{equation}
		\frac{\lambda}{2}\sum_{j = q + 2}^{k} \| \Delta^{q + 1} a_{j}\|_{0},
		\label{eq:l0}
\end{equation}
where $\|.\|_{0}$ is the L$_0$ norm, i.e. $\|x\|_{0} = 0$ if $x = 0$ and $\|x\|_{0} = 1$ otherwise, and where the parameter $\lambda > 0$ tunes the tradeoff between goodness of fit and regularity of the spline.
This penalty allows to remove a knot $t_{j^{*}}$ that is not relevant for the regression, to merge the adjacent intervals $[t_{j^{*} - 1}, t_{j^{*}})$ and $[t_{j^{*}}, t_{j^{*} + 1})$ and to continue the fitting procedure with a spline defined over the remaining knots.
		When $\lambda \to 0$, the fitted function is a B-spline with all knots $t_{1}, \dots, t_{k}$ and when $\lambda \to \infty$, the fitted function is a polynomial of degree $q$.

However, the penalty in Equation \eqref{eq:l0} is non differentiable and the estimation is therefore computationally non-tractable.
To overcome this difficulty, an approximation method for the L$_{0}$ norm is introduced in the next section.

\subsection{Adaptive ridge}
\label{section:ar}
Following the work from \cite{Rippe2012VisualizationGenomicChanges} and \cite{Frommlet2016AdaptiveRidgeProcedure}, we approximate the L$_{0}$ norm by using an iterative procedure called Adaptive Ridge.
The new objective function is the weighted penalized sum of squares:
\begin{equation}
		\mathrm{WPSS}\left( \bm{a}, \lambda \right) = \|\bm{y} - \bm{B} \bm{a}\|_{2}^{2} + \frac{\lambda}{2} \sum_{j = q + 2}^{q + k + 1} w_{j} \left( \Delta^{q + 1} a_{j} \right)^{2},
		\label{eq:wpss}
\end{equation}
where $\Delta a_{j} = a_{j} - a_{j - 1}$ is the first order difference operator, $\Delta ^{i} a_{j} = \Delta^{i - 1} \Delta a_{j}$, and $w_{j}$ are positive weights.
The penalty is close to the L$_{0}$ norm penalty when the weights are iteratively computed from the previous values of the parameter $\bm{a}$ following the formula:
\begin{equation*}
		w_{j} = \left( \left( \Delta ^{q + 1} a_{j} \right)^{2} + \varepsilon^{2} \right) ^{-1},
\end{equation*}
where $\varepsilon > 0$ is a small constant.
Indeed the function $x \mapsto x^{2}/\left(x^{2} + \varepsilon^{2}\right)$ approximates the function $x \mapsto \|x\|_{0}$ when $\varepsilon$ is sufficiently small.
In practice, one typically sets $\varepsilon = 10 ^{-5}$ \citep{Frommlet2016AdaptiveRidgeProcedure}.
At convergence, $\left(\Delta ^{q + 1} a_{j}\right)^{2} w_{j} \simeq \|\Delta^{q + 1} a_{j}\|_{0}$ gives a measure of how relevant the $j$-th knot is.
One chooses a threshold of $10^{-2}$ and selects the knots with a weighted differences higher than $0.99$, which we note $t_{j}^{\text{sel}}$.
The number of selected knots will be noted $k_{\lambda}$, such that the number of parameters of the selected spline is $q + k_{\lambda} + 1$.
Since the selected knots are present in breakpoints of the curve, one then fits unpenalized B-splines over the knots $\bm{t}^{\text{sel}}$, as explained in Section~\ref{section:b_spline_regression}.
Consequently, this method provides a regression model that is both regularizing and simple, in the sense that the model dimension is small.

We note that \cite{Frommlet2016AdaptiveRidgeProcedure} give a more general formula for the weights that allows to approximate any  L$_{p}$ norm, for $p > 0$.
In particular, the L$_{1}$ norm could be chosen, which induces both shrinkage and selection of the coefficient.
Let us note that this method was already developed by \cite{Eilers2005Quantilesmoothingarray} with B-splines of order $1$ using an exact L$_{1}$ norm and a median regression solver.

\begin{algorithm}
  \caption{Adaptive Ridge Procedure for Spline Regression
    \label{alg:aridge}}
\begin{flushleft}
		\textbf{Input:} $\bm{x}, \bm{y}, \lambda$\\
		\textbf{Output:} $\bm{\hat{a}}$
\end{flushleft}
	\vspace{-1.20cm}
  \begin{algorithmic}[1]
    \Statex
	\Function{Adaptive-Spline }{$\bm{x}, \bm{y}, \lambda$}
      \Let{$\bm{a}$}{$\bm{0};$} $\quad \bm{w} \gets \bm{1}$
	  \While{not converge}
	  \Let{$\bm{a}^{\text{new}}$}{$\arg\min_{\bm{a}} \mathrm{WPSS}\left( \bm{a}, \lambda \right)$}

	  \Let{$w_{j}$}{$\left( \left( \Delta ^{q + 1} a_{j}^{\text{new}} \right)^{2} + \varepsilon^{2} \right) ^{-1}$}
        \Let{$\bm{a}$}{$\bm{a}^{\text{new}}$}
      \EndWhile
	  \State {\bf Compute} $\bm{t}^{\text{sel}}$ using $\left(\Delta^{q + 1}\bm{a}\right)^{2}\bm{w}$
	  \Let{$\bm{\hat{a}}$}{$\arg\min_{\bm{a}} \mathrm{SS}\left( \bm{a}, \bm{t}^{\text{sel}} \right)$}
	  \State \Return{$\bm{\hat{a}}$}
    \EndFunction
  \end{algorithmic}
\end{algorithm}

$\mathrm{WPSS\left( \bm{a}, \lambda \right)}$ of Equation \eqref{eq:wpss} easily rewrites
\begin{equation}
		\| \bm{y} - \bm{B} \bm{a} \| _{2}^{2} + \lambda \bm{D}^{T} \bm{W} \bm{D} \bm{a},
		\label{eq:wpss_2}
\end{equation}
where $ \bm{W}= \text{diag}\left( \bm{w} \right)$ and $\bm{D}$ is the matrix representation of the difference operator $\Delta^{q + 1}$.
The minimization of $\mathrm{WPSS}$ is explicit:
\begin{equation}
		\bm{\hat{a}} = \left( \bm{B}^{T}\bm{B} + \lambda \bm{D}^{T} \bm{W} \bm{D} \right) ^{-1} \bm{B} ^{T}\bm{y}.
		\label{eq:nr}
\end{equation}
A detailed explanation of the adaptive ridge procedure is given in Algorithm \ref{alg:aridge}.

The penalty term is conveniently written with the circulating matrix $\bm{D}$.
However, for computational efficiency, $\bm{D}$ is never computed and instead we implement a fast computation algorithm for the penalty term.
More details about the implementation are given in Section \ref{section:implementation}.

\paragraph{Relation to P-Splines}
It is interesting to note that A-splines are closely related to P-splines \citep{Eilers1996FlexibleSmoothingBsplines}, whose objective function writes:
\begin{equation}
		\mathrm{PSS}\left( \bm{a}, \lambda \right) = \mathrm{SS}(\bm{a}) + \frac{\lambda}{2}\sum_{j = p + 1}^{k + q + 1} \left( \Delta^{p} a_{j} \right) ^{2},
		\label{eq:pss}
\end{equation}
where the difference order $p$ is a parameter to be chosen.
Thus, the implementation of A-splines can be seen as a weighted P-splines fitting.
The philosophies of A-splines and P-splines are however very different.
P-splines avoid choosing the best knots by penalizing over the differences of the coefficients.
Instead, we directly choose the best knots for spline regression.

\section{Generalized Linear Model}
\label{section:glm}

Spline regression has also been used to fit values in the general linear model setting, like in \citet{Eilers1996FlexibleSmoothingBsplines, Hastie2001ElementsStatisticalLearning}.
In this section, we extend A-spline regression to the generalized linear model.
In this setting, one estimates $\bm{\mu} = \esp{\bm{y} | \bm{x}} = g^{-1}\left( \bm{B} \bm{a} \right)$, where $g$ is the canonical link function and the variance of $\bm{y}$ is a function $V$ of $\bm{\mu}$: $\var{y} = V\left( \bm{\mu} \right)$.
Like the linear model, $\bm{\mu}$ can be estimated using spline regression.
The generalized linear model is fitted using the Iteratively Reweighted Least Squares (IRLS) algorithm \citep[][Section~2.5]{McCullagh1989GeneralizedLinearModels}.
With weighted penalization, the IRLS iteration writes:
\begin{equation}
		\bm{\hat{a}}^{\left( k + 1 \right)} = \left( \bm{B}^{T}\bm{\Omega}^{\left( k \right)}\bm{B} + \lambda \bm{D}^{T}\bm{W}\bm{D} \right)^{-1}\bm{B}^{T}\left( \bm{\Omega}^{\left( k \right)}\bm{B}\bm{\hat{a}}^{\left( k \right)} + \bm{y} - \bm{\mu}^{\left( k \right)} \right)
		\label{eq:nr_glm}
\end{equation}
where $k$ is the step index and $\bm{\Omega}^{\left( k \right)}$ is the diagonal matrix with entries
\begin{equation*}
		\omega_{i,i}^{\left( k \right)} = \frac{1}{V\left( \mu_{i}^{\left( k \right)} \right) g'\left( \mu_{i}^{\left( k \right)} \right) ^{2}},
\end{equation*}
with $\mu_{i}^{\left( k \right)} = g^{-1}\left( \bm{B}_{i}\bm{\hat{a}}^{\left( k \right)} \right)$.
In practice, the estimation procedure in Algorithm \ref{alg:aridge} remains the same, except that $\mathrm{WPSS}$ is minimized by the Newton-Raphson procedure given in Equation \eqref{eq:nr_glm}.

\section{Choice of the Penalty Constant}
\label{section:penalty}

In this section, one selects the penalty that performs the best trade-off between goodness of fit and regularity.
A first criterion is the AIC, which was used by \cite{Eilers1996FlexibleSmoothingBsplines} in a similar context:
\begin{equation}
		\mathrm{AIC}(\lambda) = \mathrm{SS}\left( \hat{\bm{a}}_{\lambda} \right)  + 2 \left( q + k_{\lambda} + 1 \right).
		\label{eq:aic}
\end{equation}

A different criterion is the Bayesian Information Criterion (BIC) \citep[see][]{Schwarz1978Estimatingdimensionmodel}:
\begin{equation}
		\mathrm{BIC}\left( \lambda \right) =  \mathrm{SS}\left( \hat{\bm{a}}_{\lambda} \right) + \left( q + k_{\lambda} + 1 \right)\log n.
		\label{eq:bic}
\end{equation}
Bayesian criteria maximize the posterior probability $\text{P}(\mathcal M_\lambda | \mathrm{data}) \propto \text{P}(\mathrm{data}|\mathcal M_\lambda) \pi(\mathcal M_\lambda)$, where $\text{P}(\mathrm{data} | \mathcal{M}_\lambda)$ is the integrated likelihood and $\pi\left( \mathcal{M}_\lambda \right)$ is the prior distribution on the model $\mathcal{M}_{\lambda}$.
This problem is equivalent to minimizing $-2 \log \text{P}(\mathcal{M}_\lambda | \mathrm{data})$. By integration 

\begin{equation*}
		\text{P}(\mathcal M_\lambda | \mathrm{data}) = \int_{\bm{a}} \text{P}(\mathrm{data}|\mathcal M_\lambda, \bm a) \pi(\bm a) d\bm{a}, 
\end{equation*}
where $\text{P}\left(\mathrm{data} | \mathcal{M}_\lambda, \bm a\right)$ is the likelihood and $\pi(\bm a)$ is the prior distribution of the parameter, which is taken constant in the following.
Thus Bayesian criteria are defined as
\begin{equation*}
		- 2 \log  \text{P}\left( \mathcal{M}_\lambda | \mathrm{data} \right) = \mathrm{SS}(\hat{\bm{a}}_{\lambda}) + \left( q + k_{\lambda} + 1 \right) \log n - 2 \log \pi(\mathcal M_\lambda) + \mathcal{O}_{\text{P}}(1).
\end{equation*}
The BIC is the Bayesian criterion obtained when one chooses a uniform prior on the model: $\pi(\mathcal M_\lambda) = 1$.
As explained by \cite{Zak-Szatkowska2011ModifiedVersionsBayesian}, a uniform prior on the model is equivalent to a binomial prior on the model dimension.
Therefore, the BIC tends to give too much importance to models of dimensions around $\frac{q + k + 1}{2}$.
Since the adaptive knot selection is performed with a large number of initial knots, this will result in underpenalized estimators.

To this effect, \cite{Chen2008ExtendedBayesianInformation} have developed an extended Bayesian information criterion called EBIC$_{0}$.
The EBIC$_0$ criterion is defined by choosing:
\begin{equation*}
		\pi(M_\lambda) = {q + k + 1 \choose q + k_{\lambda} + 1}^{-1}
\end{equation*} 
and
\begin{equation}
		\mathrm{EBIC}_0\left( \lambda \right) = \mathrm{SS} \left(\hat{\bm{a}}_{\lambda}\right) + \left( q + k_{\lambda} + 1 \right) \log n + 2 \log {q + k + 1 \choose q + k_{\lambda} + 1}.
\label{eq:ebic}
\end{equation}

The EBIC$_{0}$ assigns the same \emph{a priori} probability to all models of same dimension.
Therefore 
the EBIC$_{0}$ will tend to choose sparse models even with a high number of initial knots.
These criteria's selection performances are compared in the next section through a simulation study.
\section{Simulation Study}
\label{section:simu}

\begin{figure}
		\subfloat[Logit Function]{
				\includegraphics[width = 0.45\textwidth]{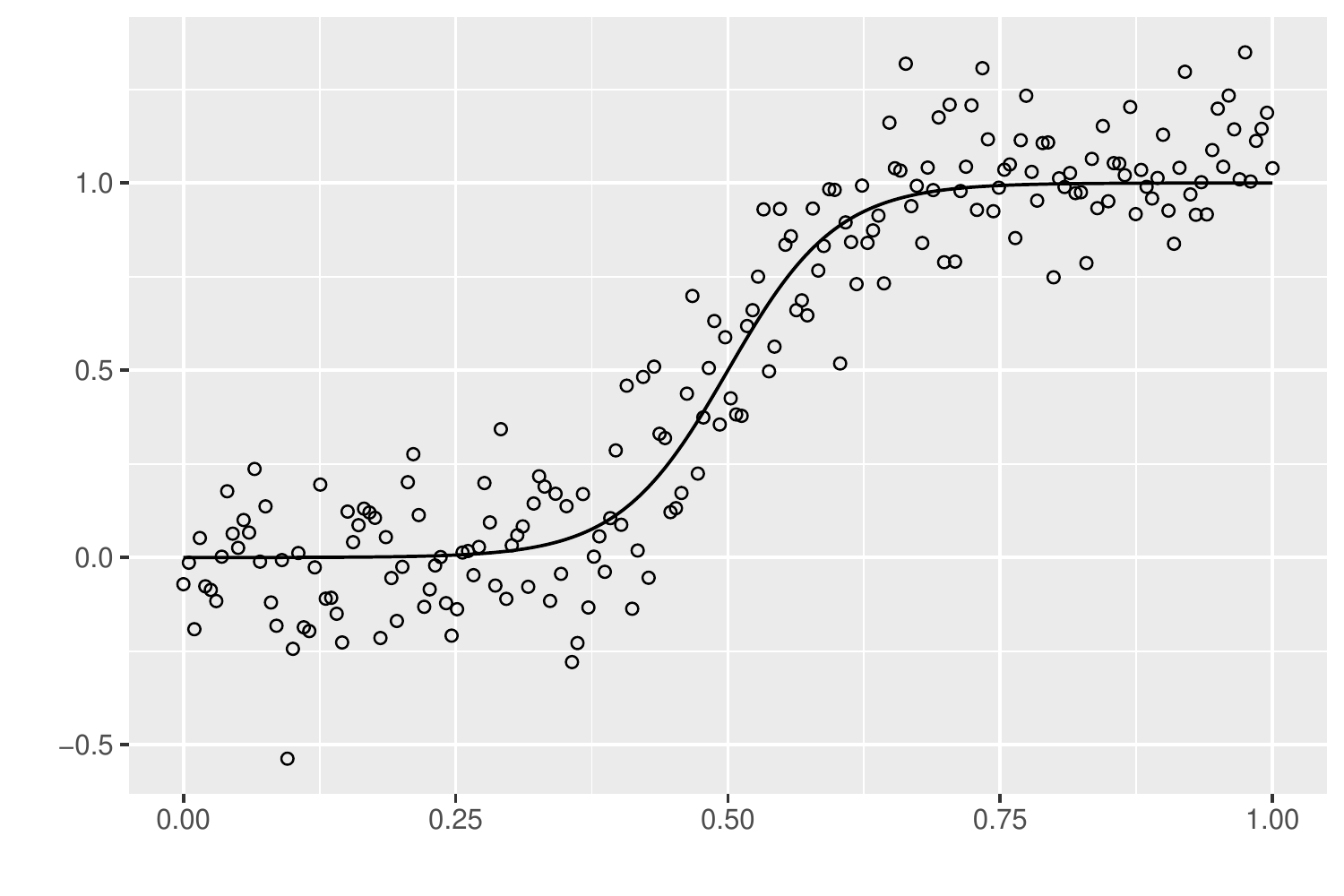}
				\label{fig:true_logit_homo}
		}
		\hfil
		\subfloat[Sine Function]{
				\includegraphics[width = 0.45\textwidth]{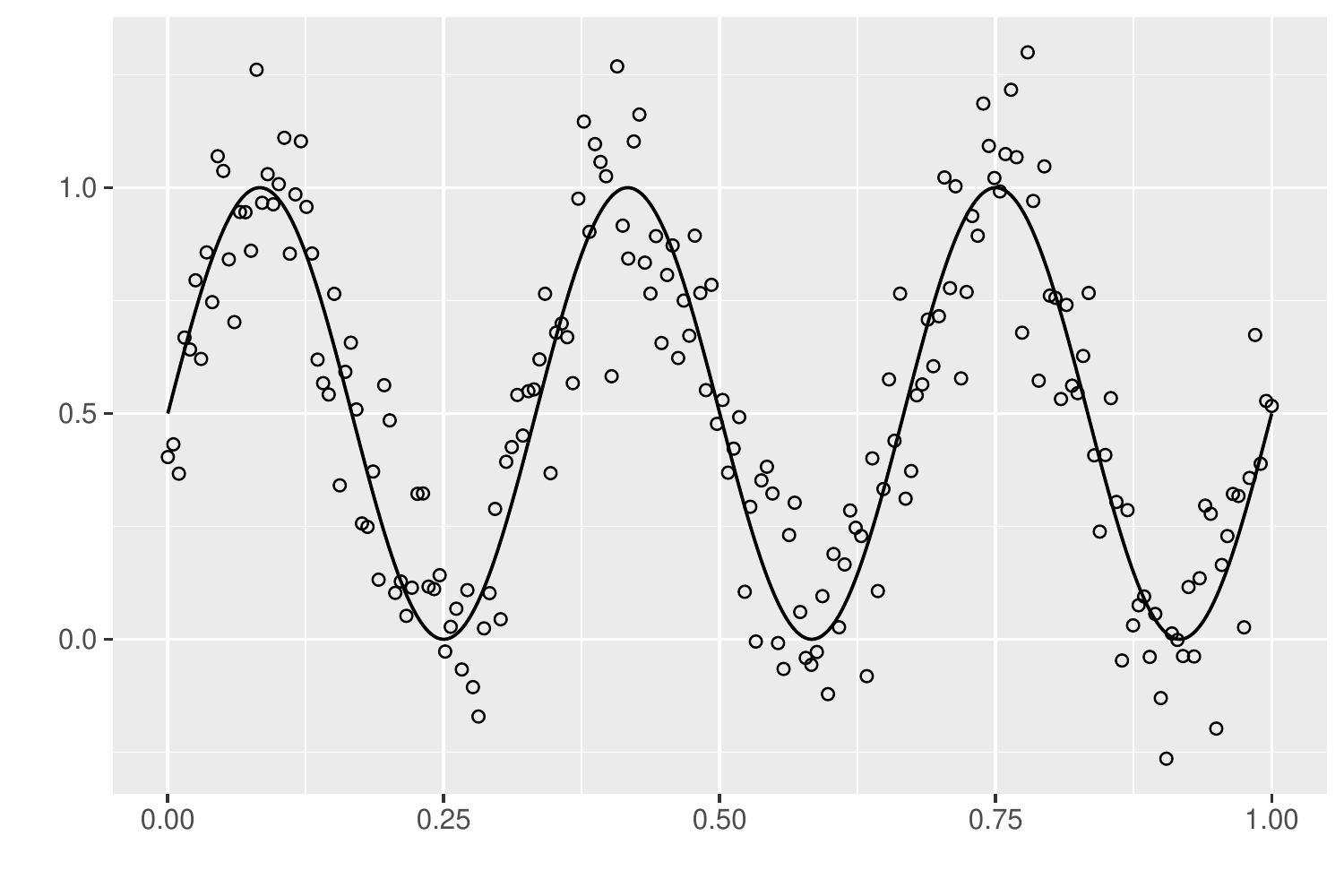}
				\label{fig:true_sine_homo}
		}
		\hfil
		\subfloat[Bump Function]{
				\includegraphics[width = 0.45\textwidth]{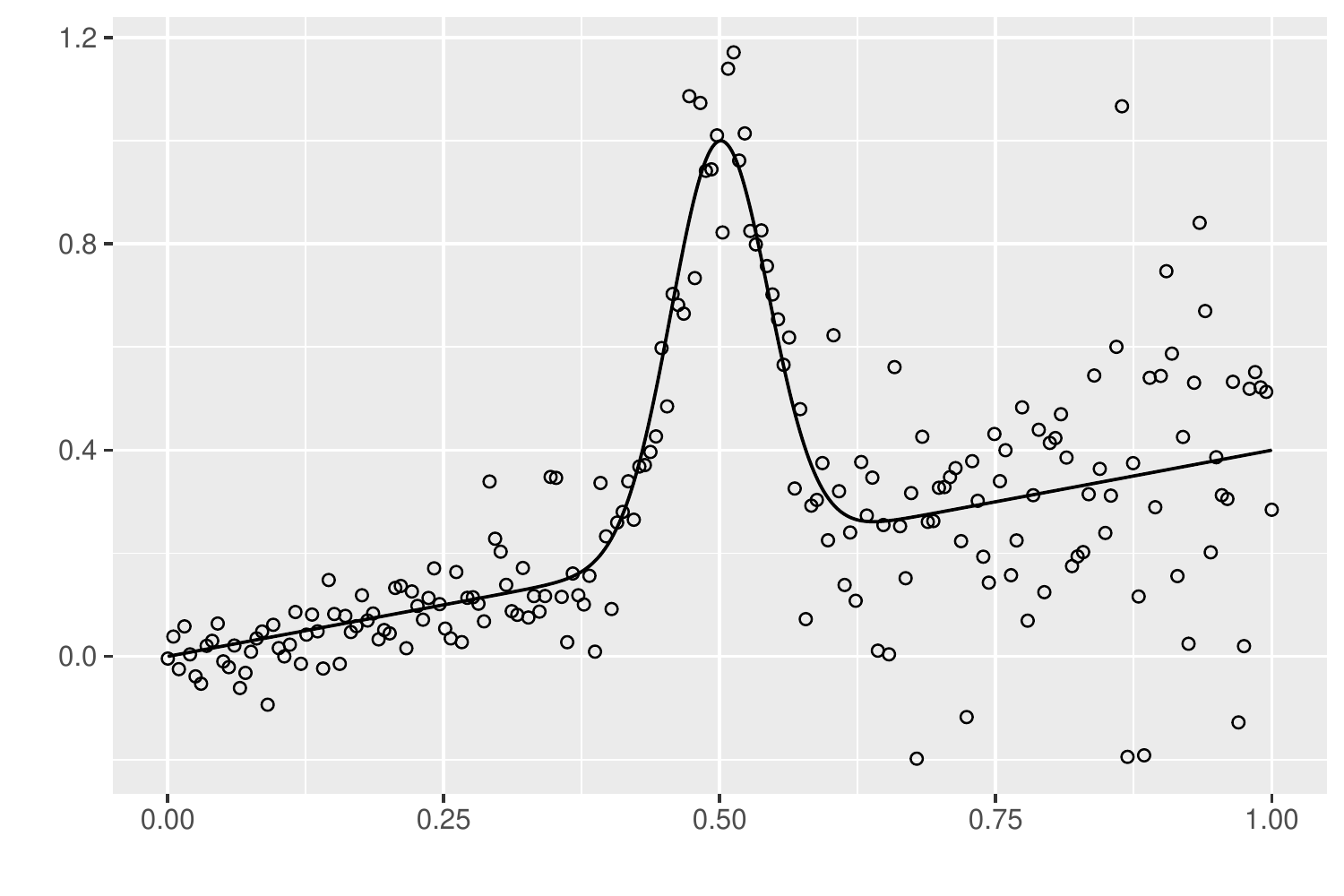}
				\label{fig:true_bump_hetero}
		}
		\hfil
		\subfloat[SpaHet Function]{
				\includegraphics[width = 0.45\textwidth]{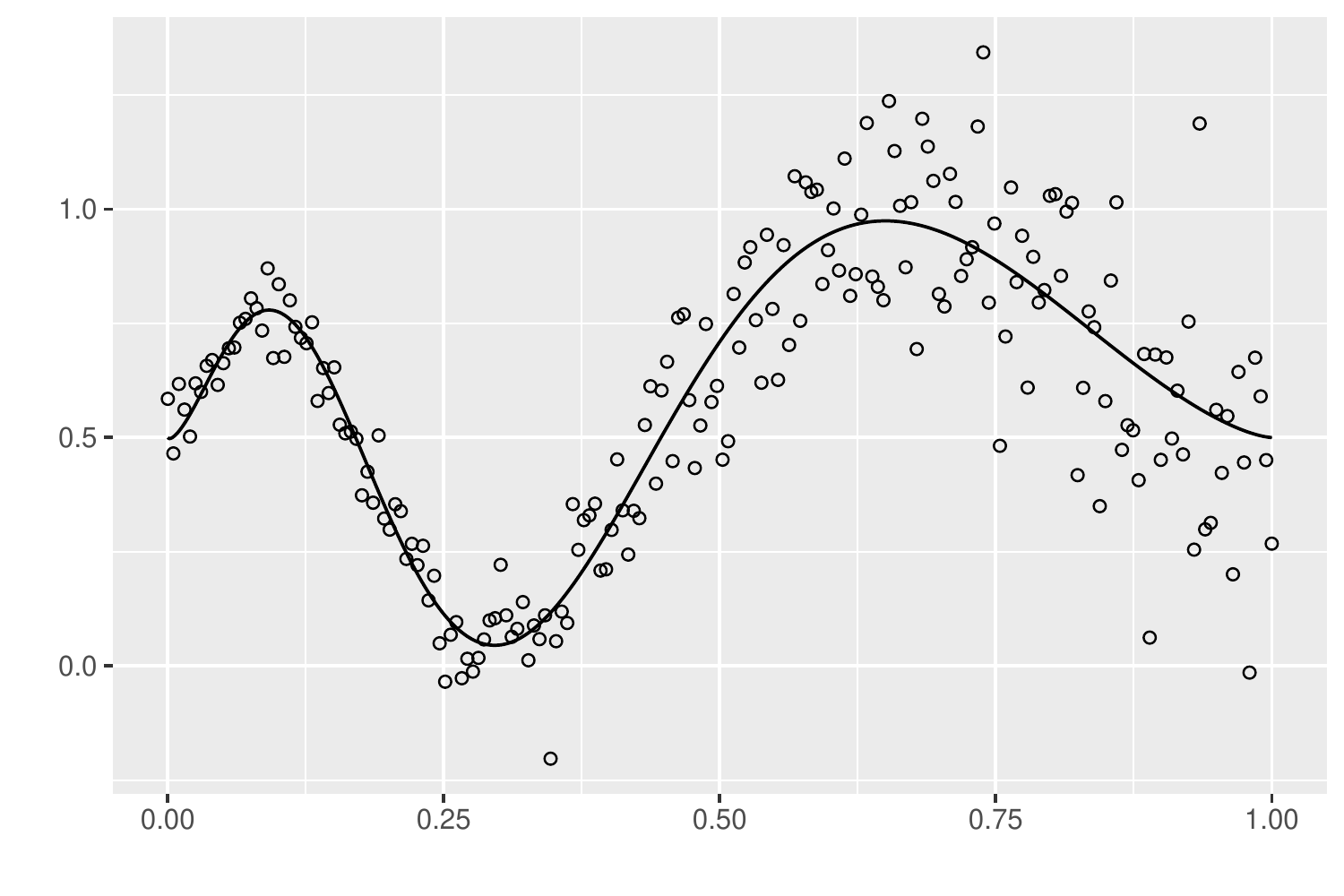}
				\label{fig:true_spa_het_hetero}
		}
		\caption{Simulated data using functions Logit (a), Sine (b), Bump (c) and SpaHet (d), in solid line. Each dataset has size $200$. The errors are chosen homoscedastic ($\sigma = 0.15$) for (a) and (b) and heteroscedastic ($\sigma_{i} = \left( 0.3 x_{i} + 0.2 \sqrt{x_{i}} \right)^{2}$) for (c) and (d).}
		\label{fig:simu_true}
\end{figure}

\begin{table}
\fontsize{10}{11}\selectfont
\centering
\subfloat[Logit Function]{
		\label{tab:criterion_logit}
\begin{tabular}{llll}
  \hline
Sample size & AIC & BIC & EBIC \\ 
  \hline
50 & 0.02220 & $\bm{0.02}$ & 0.02418 \\ 
  100 & 0.00754 & 0.00324 & $\bm{0.00248}$ \\ 
  200 & 0.00285 & 0.00136 & $\bm{0.00127}$ \\ 
  400 & 0.00131 & $\bm{0.00071}$ & 0.00072 \\ 
   \hline
\end{tabular}

}
\hfil
\subfloat[Sine Function]{
		\label{tab:criterion_sine}
\begin{tabular}{llll}
  \hline
Sample size & AIC & BIC & EBIC \\ 
  \hline
50 & 0.02239 & $\bm{0.02001}$ & 0.02459 \\ 
  100 & 0.00755 & 0.00486 & $\bm{0.00458}$ \\ 
  200 & 0.00316 & $\bm{0.00231}$ & 0.00247 \\ 
  400 & 0.00156 & $\bm{0.00132}$ & 0.00141 \\ 
   \hline
\end{tabular}

}
\hfil
\subfloat[Bump Function]{
		\label{tab:criterion_bump}
\begin{tabular}{llll}
  \hline
Sample size & AIC & BIC & EBIC \\ 
  \hline
50 & 0.02000 & $\bm{0.01801}$ & 0.02211 \\ 
  100 & 0.00735 & 0.00627 & $\bm{0.00479}$ \\ 
  200 & 0.00354 & 0.00234 & $\bm{0.00217}$ \\ 
  400 & 0.00177 & 0.00106 & $\bm{0.001}$ \\ 
   \hline
\end{tabular}

}
\hfil
\subfloat[SpaHet Function]{
		\label{tab:criterion_spa_het}
\begin{tabular}{llll}
  \hline
Sample size & AIC & BIC & EBIC \\ 
  \hline
50 & 0.02082 & $\bm{0.01784}$ & 0.02138 \\ 
  100 & 0.00727 & 0.00509 & $\bm{0.00371}$ \\ 
  200 & 0.00333 & 0.00194 & $\bm{0.00161}$ \\ 
  400 & 0.00170 & 0.00081 & $\bm{8e-04}$ \\ 
   \hline
\end{tabular}

}
\caption{Mean squared errors of adaptive spline regression for different selection criteria and for different sample sizes. Different datasets are simulated using four different functions: the Bump function (a), the Logit Function (b), the Sine function (c) and the SpaHet function (d). The smallest value of each row is highlighted in bold.}
\label{tab:criterion}
\end{table}

\begin{figure}
		\centering
		\includegraphics[width = 160 mm, height = 267 mm, keepaspectratio]{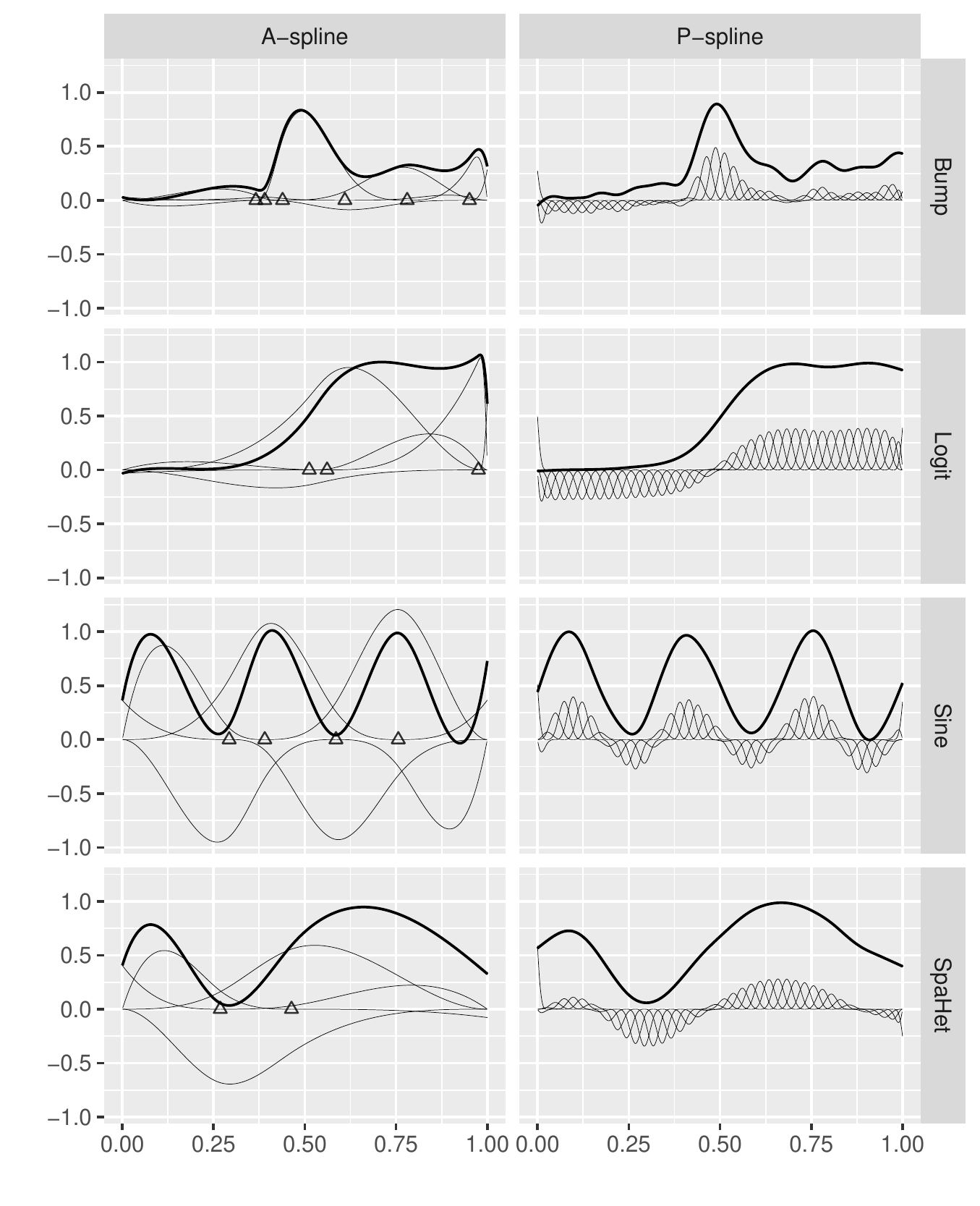}
		\caption{A-spline and P-spline regressions over different functions (tick lines). Basis decomposition of the fitted splines are represented in thin lines. For the A-spline regression, triangles represent the selected knots. The sample size is $200$.}
		\label{fig:simu_estimate}
\end{figure}

\begin{figure}
		\centering
		\subfloat[Logit]{
				\includegraphics[width = 0.45\textwidth, keepaspectratio]{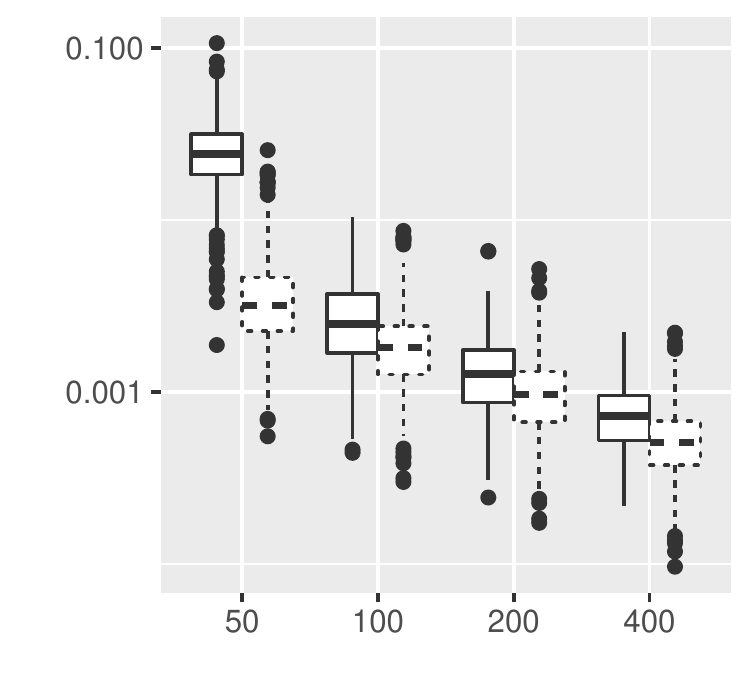}
				\label{fig:logit}
		}
		\hfil
		\subfloat[Sine]{
				\includegraphics[width = 0.45\textwidth, keepaspectratio]{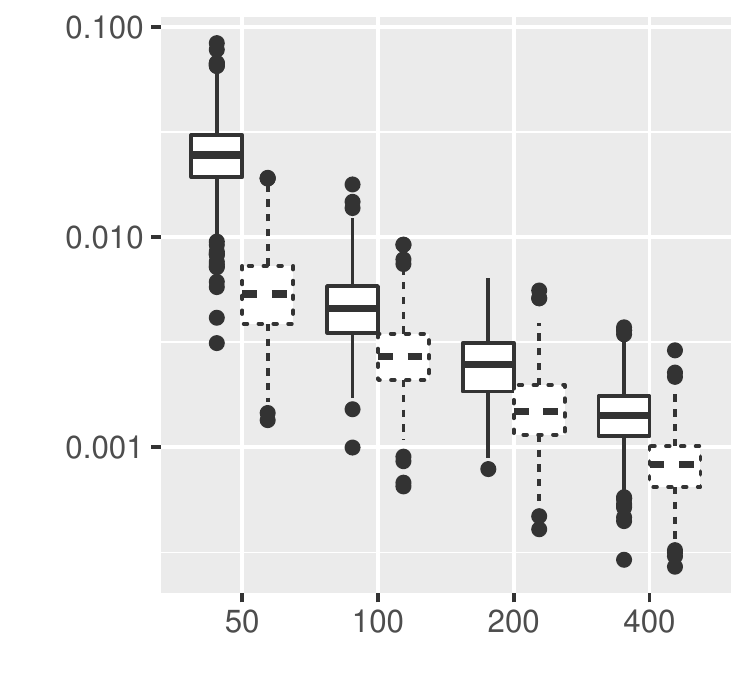}
				\label{fig:mse_sine}
		}
		\hfil
		\subfloat[Bump]{
				\includegraphics[width = 0.45\textwidth, keepaspectratio]{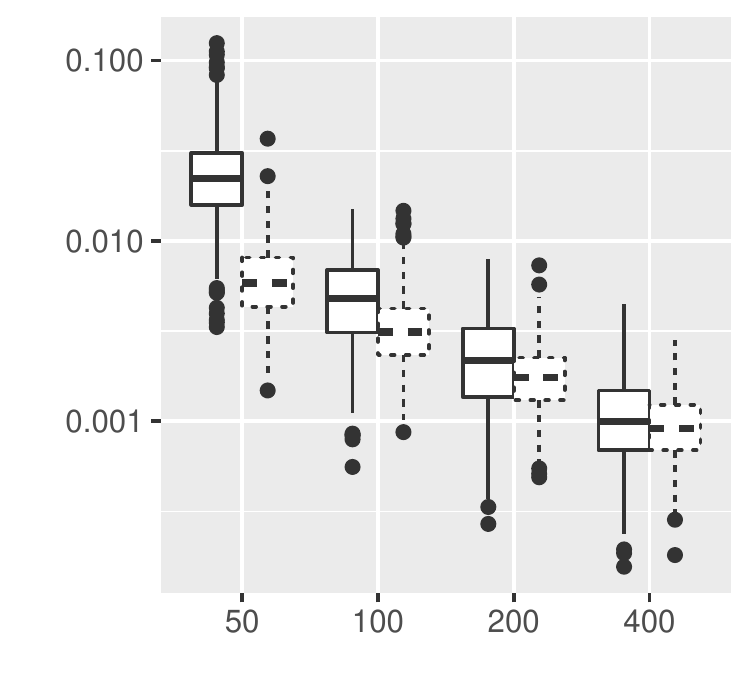}
				\label{fig:mse_bump}
		}
		\hfil
		\subfloat[SpaHet]{
				\includegraphics[width = 0.45\textwidth, keepaspectratio]{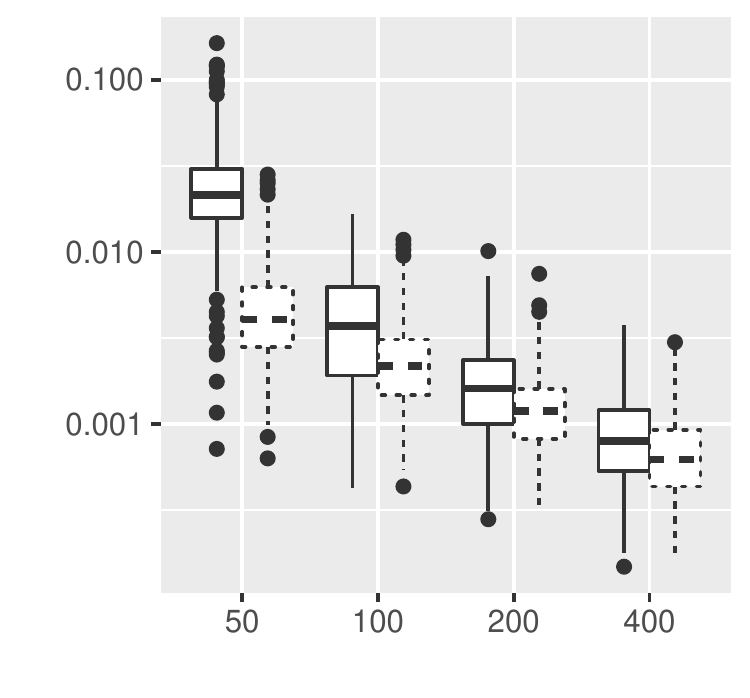}
				\label{fig:mse_spahet}
		}
		\caption{Mean squared errors of A-spline (solid line) and P-spline (dashed line) estimates for different sample sizes: $50$, $100$, $200$, and $400$. The simulations are performed with the \emph{Bump}, \emph{Logit}, \emph{Sine}, and \emph{SpaHet} functions and repeated $500$ times.}
		\label{fig:mse}
\end{figure}

\subsection{Comparing the Selection Criteria}
\label{section:simu_crit}

A simulation study has been conducted to compare the performances of the three criteria.
Data are simulated as follows.
The $x_{i}$ are taken uniformly over $[0, 1]$ and $y_{i}$ are simulated using Equation \eqref{eq:regression}, where $f$ is a known function and $\varepsilon_{i} \stackrel{}{\sim} \mathcal{N}\left( 0, \sigma_{i}^{2} \right)$.
We use four different functions: the \emph{Bump} function
\begin{equation*}
		f_{1}\left( x \right) = 0.4 \left(x + 2 \exp \left[ -\left\{ 16\left( x - 0.5 \right) \right\}^{2} \right]\right),
		\label{eq:bump}
\end{equation*}
the \emph{Logit} function
\begin{equation*}
		f_{2}\left( x \right) = \frac{1}{1 + \exp \left\{ -20 \left( x - 0.5 \right) \right\}},
		\label{eq:logit}
\end{equation*}
the \emph{Sine} function
\begin{equation*}
		f_{3}\left( x \right) = 0.5 \sin\left( 6 \pi x \right) + 0.5,
		\label{eq:sine}
\end{equation*}
and the \emph{SpaHet} -- for \emph{spatially heterogeneous} -- function
\begin{equation*}
		f_{4}\left( x \right) = \sqrt{x \left( 1 - x \right)} \sin\left( \frac{2 \pi \left( 1 + 2 ^{-3/5} \right)}{x + 2 ^{-3/5}} \right) + 0.5.
		\label{eq:spa_het}
\end{equation*}
These functions were used by \cite{Wand2000ComparisonRegressionSpline} and \cite{DavidRuppert2002SelectingNumberKnots} in similar contexts for benchmarking the efficiency of spline regression.
The functions $f_{1}$ to $f_{4}$ have been rescaled in order to vary in $\left[ 0, 1 \right]$, so that all simulation cases have similar signal-to-noise ratios.
We choose homoscedastic errors $\sigma_{i} = 0.15$ for the functions \emph{Logit} and \emph{Sine} and heteroscedastic errors for the \emph{Bump} and \emph{SpaHet} functions: $\sigma_{i} = \left( 0.3 x_{i} + 0.2 \sqrt{x_{i}} \right) ^{2}$, so that the variance increases from $0$ when $x = 0$ to $0.25$ when $x = 1$.
Data are simulated with sample sizes $50$, $100$, $200$, and $400$.
Illustration of the functions and of the simulated data are given in Figure \ref{fig:simu_true}.
For each example $500$ datasets were simulated.
A-splines are fitted and we compare the Mean Squared Error (MSE) of the estimated function for the three criteria:
\begin{equation*}
		\| f - \hat{f} \|^{2}_{2} = \int_{0}^{1} \left( f\left( x \right) - \hat{f}\left( x \right) \right)^{2} dx.
\end{equation*}

The median MSEs are displayed in Table \ref{tab:criterion} for each value of the sample size.
For all functions and for all criteria, the MSE decreases with the sample size, as is expected.
The comparison between the criteria brings the same conclusions for all four functions: the BIC and EBIC$_{0}$ always perform better than the AIC.
Moreover, note that the EBIC$_{0}$ always outperforms the BIC for the sample size $100$, and performs almost as well for the sample size $200$.
In conclusion, the BIC and EBIC$_{0}$ are to be preferred over the AIC and overall; the EBIC$_{0}$ seems a better choice than the BIC.

\subsection{Comparing A-splines with P-splines}
\label{section:simu_estimation}

In this section, the performance of A-splines is compared to penalized spline regression methods.
For the sake of simplicity, we limit our study to comparing A-splines and P-splines.
We use the same simulation setting as the previous section.
We use the EBIC$_{0}$ criterion to select the penalty.

Figure \ref{fig:simu_estimate} represents the fitted functions with A-splines and P-splines for the four functions with datasets of size $200$.
The thick lines represent the estimated functions; the thin lines represent the splines' basis decomposition.
With every function, A-spline and P-spline yield similar estimates.
The basis decomposition highlights that A-spline selects very sparse models, which are also simpler.
Over the $500$ replications, A-spline selects a median number of $9$ splines for the \emph{Bump} function, $6$ for the \emph{Logit} function, $11$ for the \emph{Sine} function, and $7$ for the \emph{SpaHet} function.

A quantitative comparison is also made to ensure that A-spline has a predictive performance comparable to P-spline.
Figure \ref{fig:mse} shows the MSE for A-splines (solid lines) and P-splines (dotted lines) for every sample size and every function.
It shows that for sample size $50$, P-splines performs better than A-splines on average.
When the sample size increases, A-splines performs almost as well as P-splines.
These two remarks are true for all four reference functions.
In conclusion, for prediction purposes P-splines are to be favored for very small dataset but for data sets of size $200$ and above, A-splines and P-splines turn out to have close to equal predictive performance.

\section{Practical Implementation}
\label{section:implementation}

In this section, the implementation of A-splines is explained in details.
Particular attention has been brought to the computation of matrix products.
Consequently, fitting A-splines is almost instantaneous: $1.3$ seconds with $q = 200$ initial knots and $n = 5000$ on a standard laptop.
In the next three sections, several bottlenecks in the computation of A-splines are addressed.
Matrix products computations are accelerated using an \texttt{Rcpp} \citep{Eddelbuettel2013SeamlessIntegrationRcpp} implementation.
An \texttt{R} implementation of the A-spline estimation procedure is publicly available in the package \texttt{aspline}\footnote{\url{github.com/goepp/aspline}}.

Let us note that the design matrix only appears in the regression model through $\bm{B}^{T}\bm{B}$ and $\bm{B}^{T}\bm{y}$, so apart from the computation of $\bm{B}$, $\bm{B}^{T}\bm{B}$, and $\bm{B}^{T}\bm{y}$, which is done only once, the algorithm does not depend on the sample size.

\subsection{Adaptive Spline Regression with Several Penalties}
\label{section:hot_start}
The penalty constant $\lambda$ tunes the tradeoff between goodness of fit and regularity.
To choose the optimal $\lambda$, regression is performed for a sequence of penalties $\bm{\lambda} = \left( \lambda_{\ell} \right), 1 \leq \ell \leq L$ and a criterion is used to determine which regression model to select.
Computing the procedure for a series of values of $\lambda$ significantly increases the computing time.
Note that a small variation of $\lambda$ yields a small variation of $\bm{\hat{a}}_{\lambda} = \arg \min _{a} \mathrm{WPSS}\left( \bm{a}, \lambda \right)$.
Consequently, $\bm{\hat{a}}_{\lambda_{\ell}}$ is a good initial point for the minimization of $\mathrm{WPSS}(\bm{a}, \lambda_{\ell+1})$.
Making use of this \emph{hot start} significantly speeds up the minimization of $\mathrm{WPSS}\left( \bm{a}, \lambda_{\ell + 1} \right)$ and thus decreases the computation time of the adaptive ridge procedure. 
This implementation of the adaptive ridge is introduced in \cite{Rippe2012VisualizationGenomicChanges} and \cite{Frommlet2016AdaptiveRidgeProcedure} and a similar idea is used in the implementation of the LASSO in the package \texttt{glmnet} \citep{Friedman2010RegularizationPathsGeneralized}.

\subsection{Fast Computation of the Weighted Penalty}
\label{section:weight_penalty}

The matrix inversion in Equations \eqref{eq:nr} and \eqref{eq:nr_glm} is the computational bottleneck of the adaptive ridge procedure.
The matrix $\bm{D}^{T} \bm{W}\bm{D}$ is symmetric and $q$-banded, and as noticed by \cite{Wand2010SemiparametricRegressionSullivan}, so is $\bm{B}^{T}\bm{B}$.
Consequently, the inversion is done using Cholesky decomposition and back-substitution, as implemented in the package \texttt{bandsolve}\footnote{\texttt{github.com/monneret/bandsolve}}.
This reduces the temporal complexity from $\mathcal{O}\left(\left( k + q + 1 \right)^{3}\right)$ to $\mathcal{O}\left( \left( k + q + 1 \right)\left( q + 2 \right) \right)$.
For example, if $k = 50$ and $q = 3$, the computation time will be reduced by a factor $500$.
It is important to note that the matrices $\bm{W}$ and $\bm{D}$ are not stored in memory: only the vector $\bm{w}$ and the first row of $\bm{D}$ are used.
This leads to improvements in spatial complexity, the details of which are not given here.

\subsection{Fast Computation of the Weighted Design Matrix}
\label{section:weight_design}
In the setting of generalized linear regression, the matrix product $\bm{B}^{T}\bm{\Omega}\bm{B}$ in Equation \eqref{eq:nr_glm} is computed at each iteration of the Newton-Raphson procedure.
Since the design matrix has $n$ rows, this operation makes the generalized linear regression computationally expensive for large datasets.
Fortunately $\bm{B}$ is sparse: it has $q + 1$ non-zero elements in each row.
Due to this structure, the product $\bm{B}^{T}\bm{\Omega}\bm{B}$ only has $\left( q + k + 1 \right)\left( q + 1 \right)$ non-zero entries.
Each entry takes $\mathcal{O}\left( \frac{n}{k} \right)$ operations to compute on average.
Thus the matrix product can be computed with a $\mathcal{O}\left(\left( q + k + 1 \right)\left( q + 1 \right) n / k\right)$ temporal complexity, compared to the $\mathcal{O}\left( \left( q + k + 1 \right)^{2}n \right)$ complexity of the naive implementation.
For instance, even with $q = 3$ and $k = 50$, this implementation is faster by a factor $\sim 700$.

\section{Real Data Applications}
\label{section:real_data}

\begin{figure}
		\centering
		\subfloat[A-splines]{
				\includegraphics[width = 0.80\textwidth]{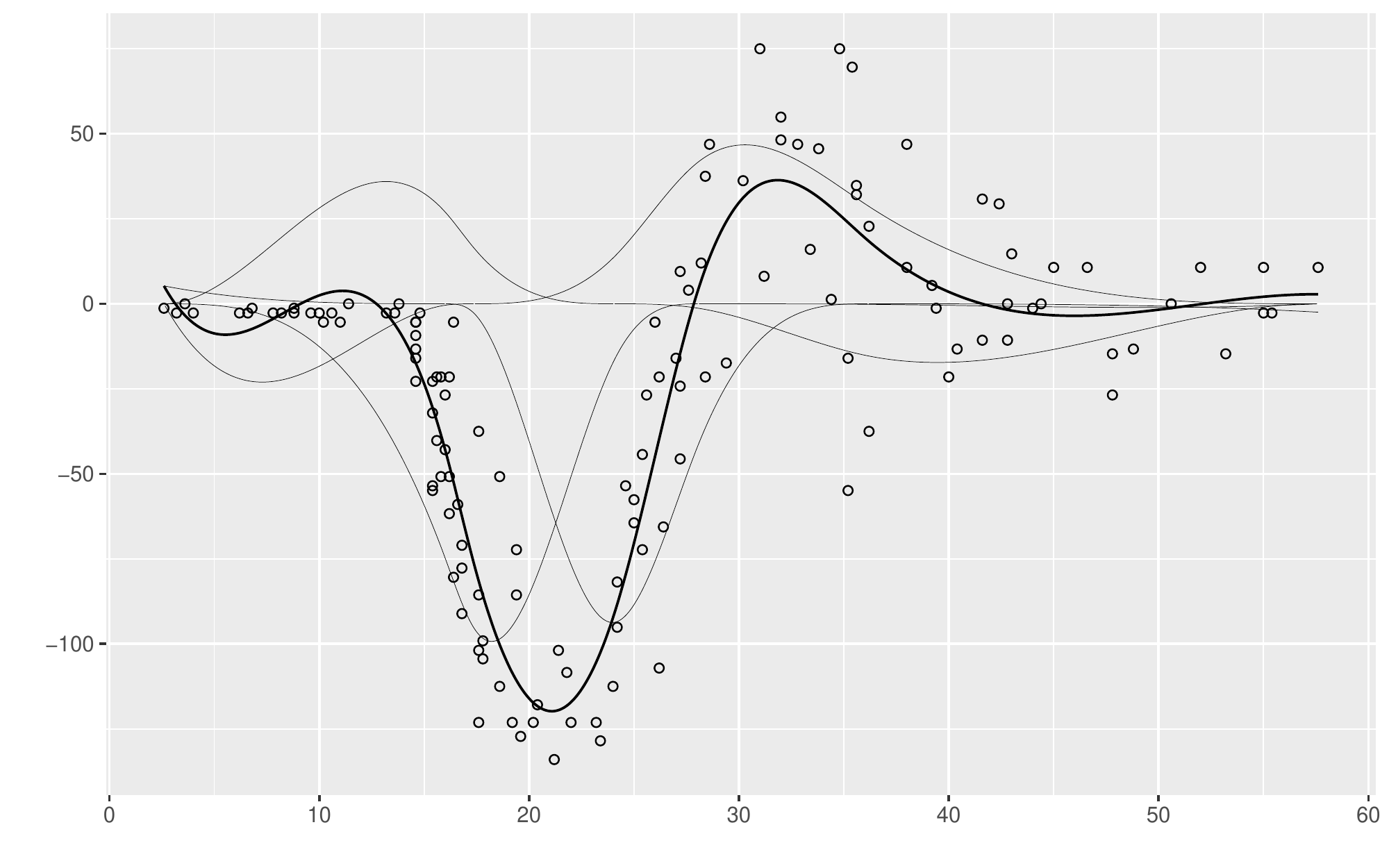}
				\label{fig:helmet_a_spline}
		}\\
		\subfloat[P-splines]{
				\includegraphics[width = 0.80\textwidth]{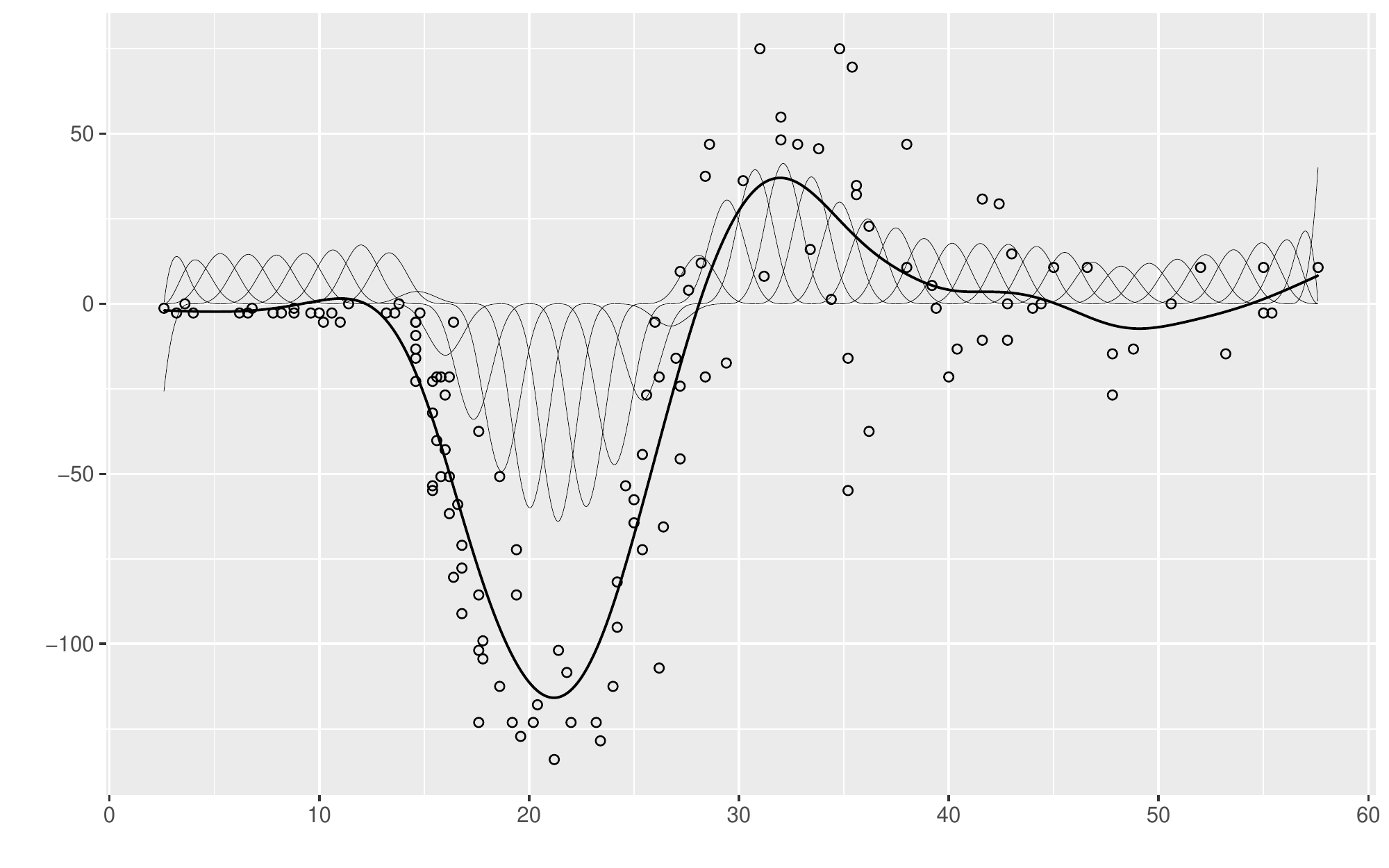}
				\label{fig:helmet_p_spline}
		}
		\caption{Motorcycle crash data: helmet acceleration (unit of \emph{g}) as a function of time (in \emph{ms}). A-spline (a) regression and P-spline (b) regression are fitted. Bold lines represent the estimates and grey lines represent the decomposition of the estimates onto the B-spline bases.}
		\label{fig:helmet}
\end{figure}
\begin{figure}
		\centering
		\includegraphics[width = 0.85 \textwidth, keepaspectratio]{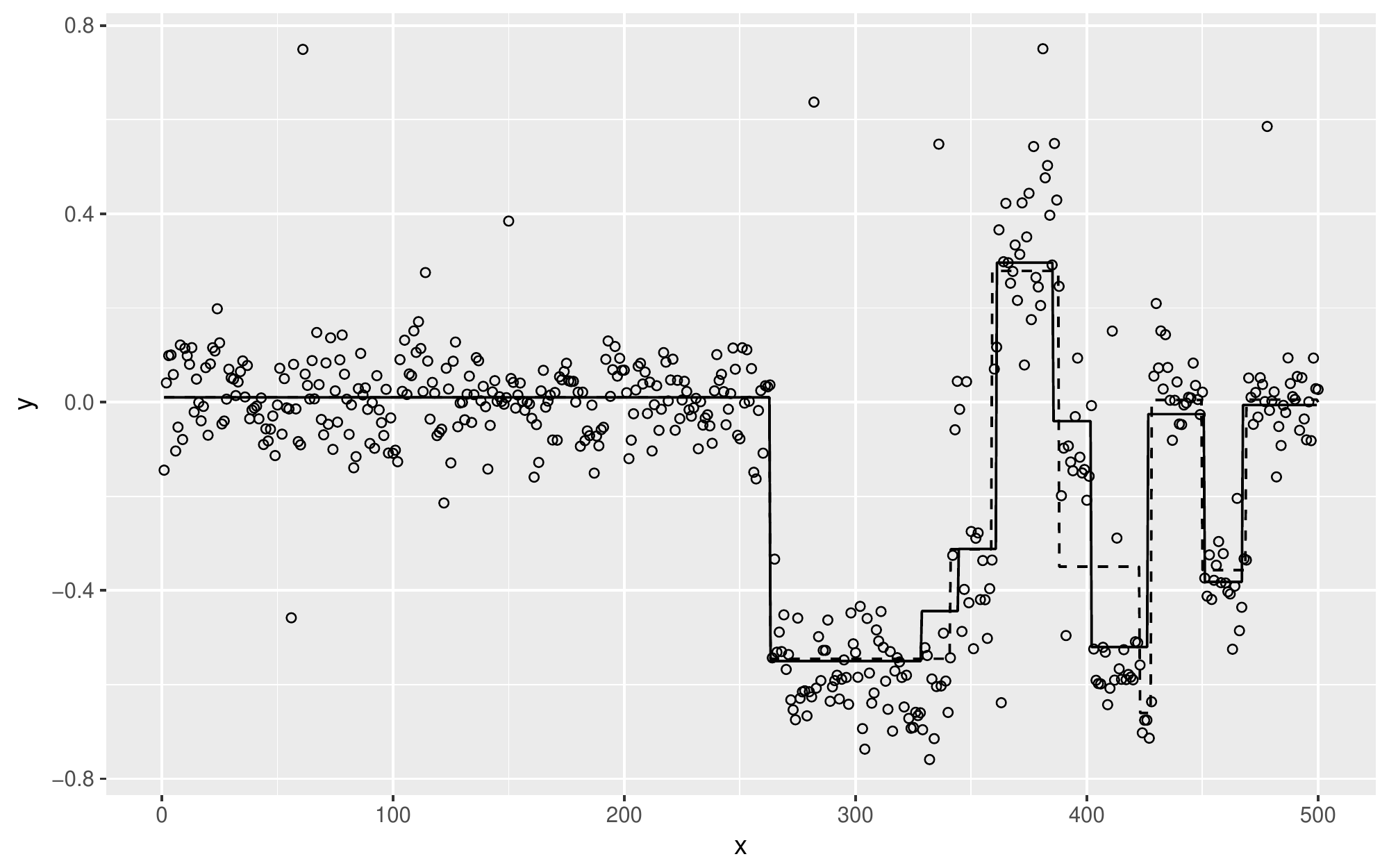}
		\caption{aCGH data of bladder cancer: probes $1$ through $500$. A-splines of order $0$ are fitted (solid line) as well as the mean values fitted using the PELT changepoint detection method (dashed line).}
		\label{fig:bladder}
\end{figure}
\begin{figure}
		\centering
		\includegraphics[width = 0.85 \textwidth, keepaspectratio]{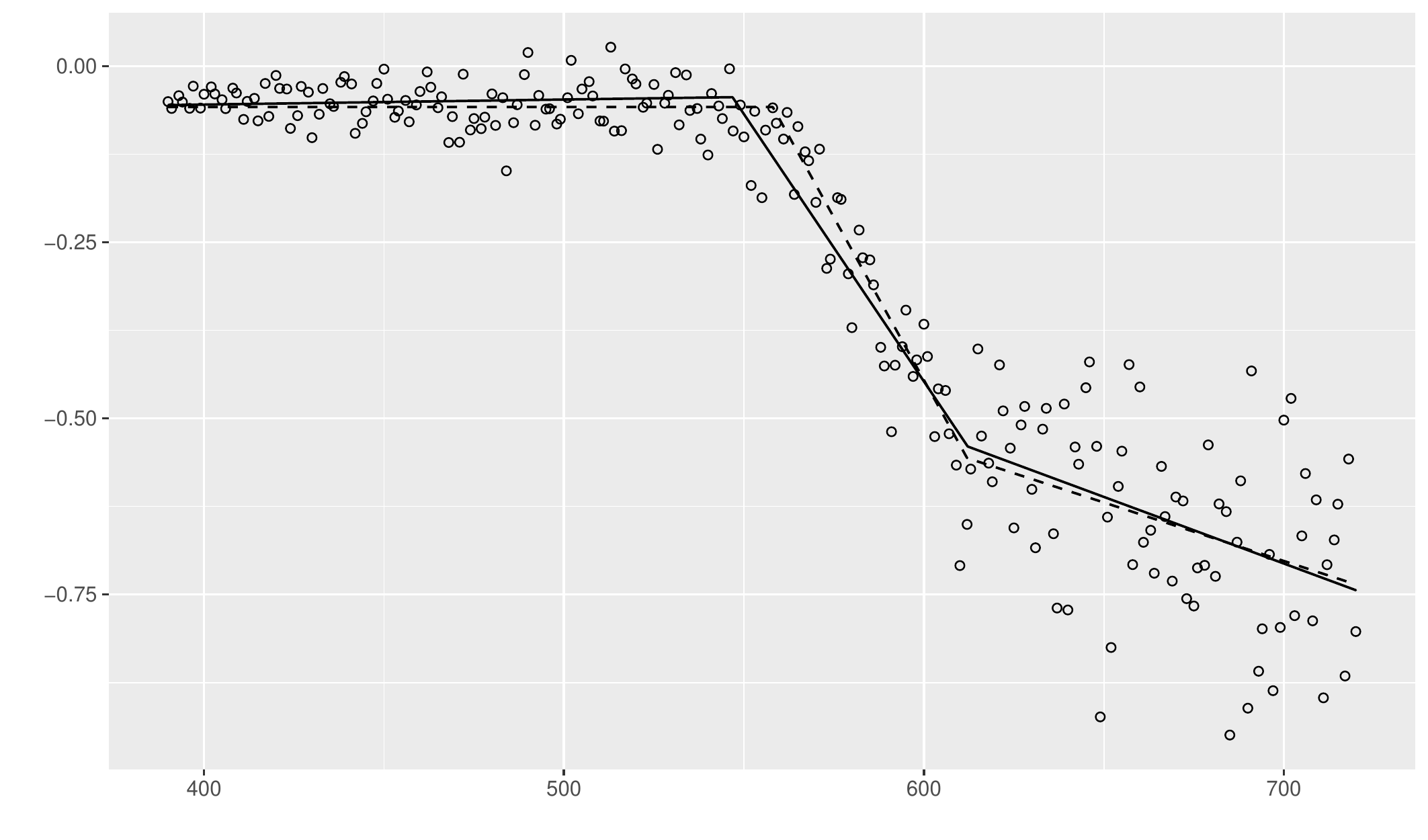}
		\caption{LIDAR data: log-ratio of light intensity as a function of the travelled distance. A-splines of order $1$ (solid line) and Multivariate Adaptive Regression Splines (dashed lines) are fitted.}
		\label{fig:lidar}
\end{figure}
\begin{figure}
		\centering
		\includegraphics[width = \figwidth mm, height = \figheight mm, keepaspectratio]{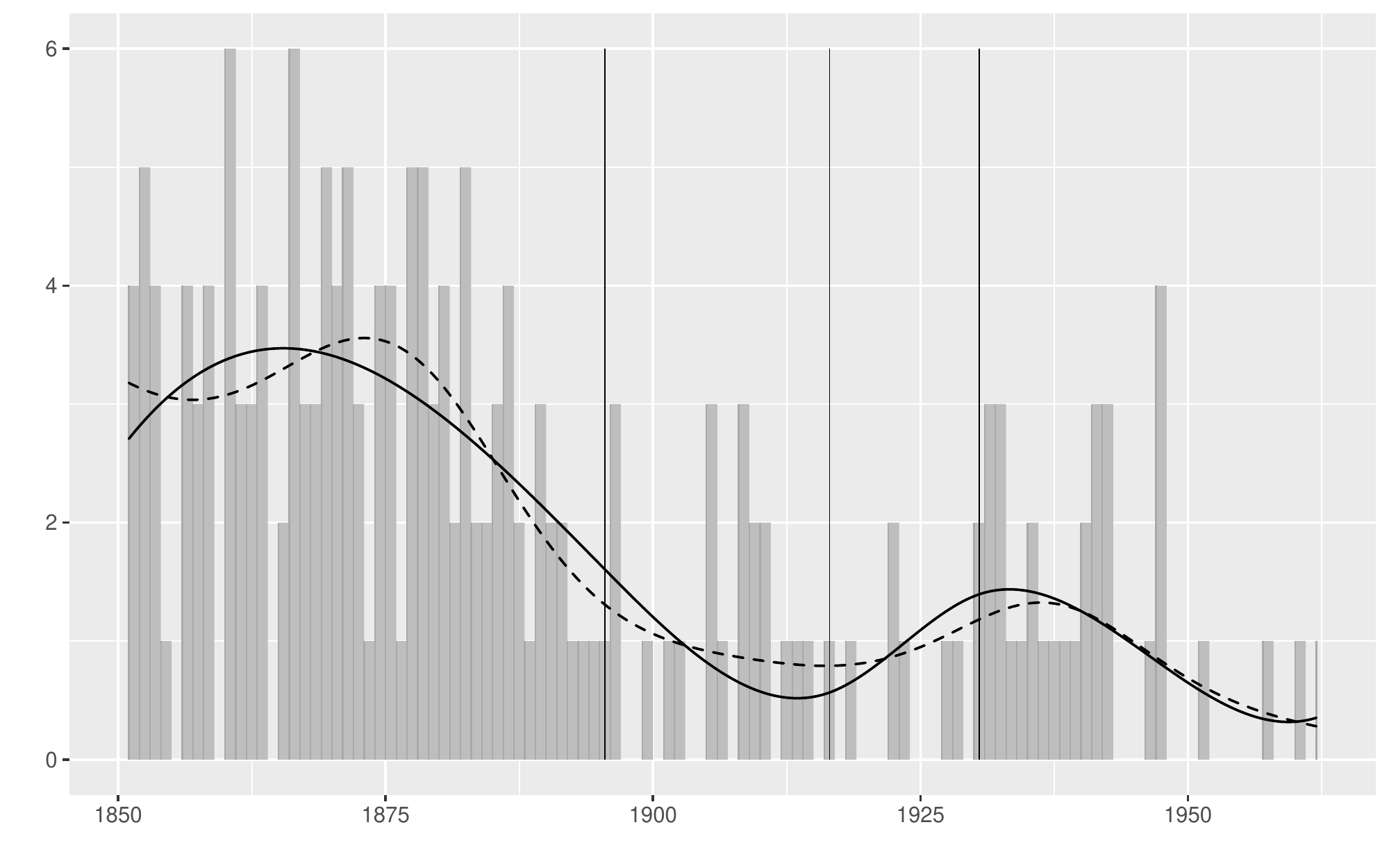}
		\caption{Yearly number of coal accidents in Britain (grey bars) with P-splines regression (dashed curve) A-spline regression (solid curve). The three knots selected by A-splines are represented by vertical lines.}
		\label{fig:coal}
\end{figure}

Our method is illustrated with several real data applications.

We first present a dataset of simulated motorcycle accidents used to crash-test helmets.
The data consists of $132$ observations of helmet acceleration (in units of $g$) measured along time after impact (in \emph{milliseconds}).
These data have being used as illustration of spline regression by \cite{Silverman1985AspectsSplineSmoothing} and \cite{Eilers1996FlexibleSmoothingBsplines} and are available in \cite{Hand1993HandbookSmallData}.
This dataset represents a good test for non parametric regression since the variance of the errors varies a great deal and there are several breakdown moments in the data.
A-spline regression of order $q = 3$ is performed (Figure \ref{fig:helmet}a).
For the sake of the illustration, our regression in compared to P-splines of order $q = 3$ (Figure \ref{fig:helmet}b).
We have set $k = 40$ equally spaced initial knots for both regression methods.
In both figures, the solid lines represent the estimated fit and the dashed lines represent the decomposition of the fit onto the B-spline family.
The two estimations are almost equal.
A-spline regression has selected only $5$ knots as relevant, and thus the fitted function is a linear combination of $5 + 3 + 1 = 9$ splines.

The second illustrative example uses a dataset of array Comparative Genomic Hybridization (aCGH) profiles for 57 bladder tumor samples \citep[see][for references and access to the data]{Stransky2006Regionalcopynumber}.
This dataset was used by \cite{Bleakley2011GroupFusedLasso} in the similar context of changepoint detection. 
The data represent the log-ratio of DNA quantity along $2215$ probes.
For the illustration, the $500$ first observations of individual $1$'s aCGH profile are used.
We fit a spline of order $0$, i.e. a piecewise constant function.
Indeed, A-splines of order $0$ perform a regression with changepoint detection of the data, which is a desired goal for these data.
The fitted spline is represented in solid line in Figure \ref{fig:bladder}.
The estimated function performs a satisfying estimation of the changepoints and of the mean values over each interval.
Our regression method estimated $9$ changepoints, each corresponding to a shift in the mean value of the signal.
Our method is compared to a popular changepoint detection algorithm (dashed line of Figure \ref{fig:bladder}) called PELT \citep{Killick2012OptimalDetectionChangepoints}.
We used \texttt{R} package \texttt{changepoint.np} \citep{Haynes2016changepointnpMethods} .
This method detects $8$ changepoints, all of which correspond to a changepoint detected by the A-spline regression.

The third example is based on the LIDAR data \citep{Sigrist1994Airmonitoringspectroscopic, Holst1996LOCALLYWEIGHTEDLEAST}, which is used by \cite{Ruppert2003SemiparametricRegression} to illustrate regression methods.
The data come from a light detection and ranging (LIDAR) experiment.
It consists of $221$ observations of log-ratio of measured light intensity between two sources, as a function of the distance travelled by the light before being reflected (in \emph{meters}).
The data are available in the \texttt{R} package \texttt{SemiPar} and are represented in Figure \ref{fig:lidar}.
The scatter plot clearly displays a smooth decrease of the y-variable.
More precisely, the y-variable is slightly decreasing for lower values of $x$. 
There is a clear decrease of the slope between $x = 550$ m and $x = 600$ m, after which the slope gradually increases.
To highlight these shifts in slope, splines of order $1$ (i.e. piecewise linear functions) are chosen to fit the data.
The A-spline fit displays two slope changes, at $x = 567$ m and $x = 607$ m.
These moments visually correspond to the two biggest shifts in slope.
We also fit \cite{Friedman1991MultivariateAdaptiveRegression}'s MARS procedure (in dashed line, Figure \ref{fig:lidar}) and compare it to A-splines.
We use an implementation of the procedure in the \texttt{R} package \texttt{earth}.
This method also selects two breakpoints of the slope, at $x = 558$ and $x = 612$, which are very close to the breakpoints detected by A-splines.


The last example uses the data of the registered number of disasters in British coal mines per year between the years $1850$ and $1962$ \citep[][]{Diggle1988EquivalenceSmoothingParameter}.
The number of coal disasters in each year is assumed to be Poisson distributed and the mean of the distribution is fitted using a Poisson regression.
The data are fitted using A-spline regression of order $3$.
The fitted curve $\hat{\bm{\mu}} = g^{-1}\left( \bm{B} \hat{\bm{a}} \right)$  is given in Figure \ref{fig:coal}.
The $3$ selected knots are represented by vertical dashed lines.
The regression is compared to P-splines (in dashed lines), which yields a similar estimation -- although less regularized.

\section{Conclusion}
In this paper we introduce a method called A-spline (for adaptive spline) performing spline regression
which automatically selects the number and position of the knots.
For that purpose, we set a large number of initial knots and use an iterative penalized likelihood approach (the adaptive ridge) to sequentially remove the unnecessary knots.
The model achieving the best bias-variance tradeoff is selected using a Bayesian criterion: either the BIC or the EBIC$_{0}$.

Our method yields sparse models which are more interpretable than classical penalized spline regressions (e.g. P-splines).
Yet, a simulation study
shows that our method has predictive performances comparable to P-splines.

When using A-spline with low order splines (e.g. 0 or 1), the approach allows performing changepoint detection. 
Indeed, A-spline of order 0 fit a piecewise constant function to the data and hence detect changepoint in terms of mean. A-spline of order 1 fits a piecewise linear continuous function (i.e. a continuous broken line) that detects changepoints in terms of slope.

A fast implementation of A-spline is provided in \texttt{R} and \texttt{Rcpp}.
Thanks to this, the computation of A-spline is very fast ($\sim 1 \text{ sec}$ for $n \sim 10000$ $k \sim 1000$ on the standard laptop), even when fitting generalized linear models with large sample sizes.

Our work can be naturally generalized to multivariate data using multidimensional B-splines.
Moreover, we limited our work to using B-splines for the sake of simplicity.
But a variety of other splines can be used instead.
For example M-splines, which are a basis of non-negative splines, could be used for fitting non-negative functions (e.g. densities)
and I-splines, which are a basis of monotonous splines, would yield a sparse isotonic regression model.
Finally, our spline regression method can be used for non-parametric transformation of variables.
In particular, splines of order $0$ could provide an automatic categorization of continuous covariates variables in regression models.



\bibliographystyle{agsm}
\bibliography{biblio.bib}
\end{document}